\documentclass[]{aa}
\pdfoutput=1
\usepackage{graphicx}
\usepackage{txfonts}
\usepackage{lscape}
\usepackage{rotating}
\graphicspath{{./figures_astroph/}}
\usepackage{natbib}
\usepackage{aalongtable}

\def\etal{et al. }

\def\Ia{SN~Ia~}
\def\Iae{SNe~Ia~}

\newcommand{\Yp}{{\tt Y}}
\newcommand{\Xp}{{\tt X}}

\begin{document}
\def\lesim{\stackrel{<}{{}_{\sim}}} 
\def\gesim{\stackrel{>}{{}_{\sim}}}
\title{PHotometry Assisted Spectral Extraction (PHASE) and identification of SNLS supernovae\thanks{Based on observations obtained with FORS1 at the Very Large Telescope on the Cerro Paranal, operated by the European Southern Observatory, Chile (ESO Large Programmes 171.A-0486 and 176.A-0589)}}
\author{S. Baumont\inst{1}, C.  Balland\inst{1,2},  
P. Astier\inst{1}, J.  Guy\inst{1}, D. Hardin\inst{1}, D. A. Howell\inst{3}, C. Lidman\inst{4}, M. Mouchet\inst{5,6}, R. Pain\inst{1}, N.  Regnault\inst{1}}

\institute{LPNHE, CNRS-IN2P3 and Universities of Paris 6 \& 7,
F-75252 Paris Cedex 05, France 
\and University Paris 11, Orsay, F-91405, France
\and Department of Physics and Astronomy, University of Toronto, 50 St. George Street, Toronto, ON M5S 3H4, Canada
\and European Southern Observatory, Alonso de Cordova 3107, Vitacura, Casilla 19001, Santiago 19, Chile
\and APC, UMR 7164 CNRS, 10 rue Alice Domon et L\'eonie Duquet, F-75205, Paris Cedex 13, France
\and LUTH, UMR 8102 CNRS, Observatoire de Paris, Section de Meudon, F-92195 Meudon Cedex, France
}
 % end \institute

\offprints{balland@lpnhe.in2p3.fr}

\date{Received; accepted} \titlerunning{PHASE extraction and SALT2 identifications}
\authorrunning{Baumont, Balland \etal}

\abstract
  {}
  %Aim
  { We present new extraction and identification techniques for supernova (SN) spectra developed within the Supernova Legacy Survey (SNLS) collaboration.}
  %Method
  {The new spectral extraction method takes full advantage of photometric information from the Canada-France-Hawa\"{\i} telescope (CFHT) discovery and reference images by tracing the exact position of the supernova and the host signals on the spectrogram. When present, 
the host 
spatial profile is measured on deep multi-band reference images and is
used to model the host contribution to the full (supernova + host) signal. The supernova is modelled as a Gaussian function of width equal to the seeing. 
A $\chi^2$ minimisation provides the flux of each component in each pixel of the 2D spectrogram.
For a host-supernova separation greater than $\gesim$ 1 pixel, the two components are recovered separately and we do not use a spectral template in contrast to more standard analyses. This new procedure permits a clean extraction of the supernova separately from the host in about 70\% of the 3rd year ESO/VLT spectra of the SNLS. A new supernova identification method is also proposed. It uses the SALT2 spectrophotometric template to combine the photometric and spectral data. A galaxy template is allowed for spectra for which a separate extraction of the supernova and the host was not possible.}
  %Result
  {These new techniques have been tested against more standard extraction and identification procedures. They permit a secure type and redshift
determination in about 80\% of cases. The present paper illustrates their performances on a few sample spectra.} 
  {}

\maketitle

\section{Introduction}

Observing Type Ia supernovae (SNe~Ia) for the purpose of constraining cosmological parameters is now a mature activity.
Large-scale projects of detection and spectrophotometric follow up of hundreds of \Iae from ground-based telescopes reach completion
today \citep{Astier06,Wood-Vasey07}, 
and future \Iae surveys will bring this number up to several thousand.
These current and future surveys yield and will yield new samples of \Iae observations of unprecedented number and quality. Extracting the most possible out of these exceptional sets of \Iae data is a challenge for the teams involved in these projects.

At redshifts greater than $z\gesim 0.2$, the problem of spectroscopically
identifying supernovae is more challenging than it is at low redshift. At
low redshift, galaxies have a larger angular size, so at the position of the
supernova (SN) the spectrograph slit usually contains little contaminating
host galaxy light. As redshift increases, apparent galaxy sizes decrease
and the galaxy light can be comparable to, or dominate, the supernova 
light at a given position. For observations of high-redshift supernovae,
new techniques are therefore 
required to efficiently recover the supernova signal (see e.g., Blondin at al. 2005).

Currently, the \Iae spectroscopic sample of the Supernova Legacy Survey (SNLS) represents more than 500 spectra taken from 8-10m diameter telescopes, both in the Southern and Northern hemispheres (VLT, Gemini N and S, Keck-I and -II) during large observing programmes of hundreds of hours run from 2003 up to the present day.  The spectra obtained with such programmes sample a large fraction of the wavelength space (depending on the instrument used) in the visible. They have restframe phases\footnote{Throughout this paper, the phase $\phi$ is the restframe age of the supernova in days with respect to the B-band maximum light.} comprised between -15 and +30 days and sufficient signal-to-noise ratios (S/N) for an unambiguous identification in $\sim 80$\% 
of cases. 

Besides their use for cosmological purposes -- that is identification of the supernova type and determination of their redshift for their inclusion in a Hubble diagram, these spectra contain valuable information to understand the nature and the diversity of SNe~Ia and test the adequacy of using them as ``standardisable", if not standard, candles. As an example, \citet{Guy07} have implemented a number of SNLS spectra in the training set of the spectrophotometric template SALT2, one of the fitters used to fit the light curves of \Iae in the 3rd year SNLS Hubble diagram. 
Moreover, comparison of the spectral properties of these high-redshift ($<z>\sim 0.5-0.6$) supernovae with their low-$z$ counterparts (from, e.g.,  the spectral time series currently collected by the SN Factory experiment, see \citealt{Aldering02} and \citealt{Matheson08}) should help in understanding possible evolutionary effects in \Iae populations \citep{Bronder08,Balland06,Balland07,Blondin06,Garavini07,Foley08}. Another interesting particularity of the SNLS spectral sample is that, due to the high average redshift of the SNLS SNe~Ia, the ultraviolet (UV) part of the supernovae spectra down to $\approx 3200$ $\AA$ is accessible for a large number of supernovae at various phases (see, e.g., \citealt{Ellis08} for a study of the UV features of 36 SNLS-Keck spectra). This is of particular interest as it has been suggested that possible evolution with redshift, due to different progenitor metallicities at low and high redshift, might be imprinted in the spectral features of this region of the spectrum \citep{Hoeflich98, Lentz00}. 
Optimal extraction of SNLS spectra and their identification are thus not only a crucial step in building the Hubble diagram but are also crucial for their subsequent use in studies of the physics of SNe~Ia. 

Standard analyses of spectra are often based on the Horne's extraction method \citep{Horne86} that uses the spatial profile of the source light to perform an optimal (minimum-variance unbiased) spectral extraction.
A drawback of this approach is that only mild geometrical distortions due, for example, to flexure of the instrument are corrected for (however, see \citet{Marsh89} for an extension of Horne's method suited for highly distorted spectra).
Another major problem of this approach is that it does not permit the separate extraction of the supernova from the host signal, except if the two components (SN and host galaxy) are well separated on the spectrogram (and in this case, correlations between the host and the SN spectra are usually unavoidable, even if the extraction window is adjusted to maximise the supernova signal and minimise the host contamination). 
In most cases, the SN/host separation is performed {\it a posteriori} by fitting 
a two-component model (built from a supernova and a galaxy template), the contribution of each component to the total model being evaluated by a $\chi^2$ minimisation \citep{Howell02,Howell05,Balland06,Balland07}. As the model is built from a (limited) set of templates, the SN and host spectra obtained by this procedure are strongly model-dependent and provide at best a hint of the host contribution to the full spectral {\it model}. \citet{Ellis08} have developed an improved separation technique based on fitting synthetic galaxy spectra to the host photometry measured
on reference images. By doing so, they still rely on a spectral model to estimate the host contribution to the full signal. \citet{Blondin05} use a convolution technique to simultaneously recover a host-free PSF-like SN 
component and its background galaxy spectrum. This method does not
use any spectral template for the host modeling and thus improves over
more classical extraction techniques. However, this technique is sensitive to
the width of the  spatial resolution {\em Gaussian} kernel used to recover the
background (host) component. The width has to be tuned by the user 
and highly depends on the spatial extension of the host galaxy. Moreover, the
method does not use photometric information that potentially goes along
with the spectrum.

To circumvent these drawbacks and improve over standard extraction methods, we have developed a dedicated reduction and extraction pipeline, called PHASE (PHotometry Assisted Spectral Extraction). This new technique uses photometric priors from Canada-France-Hawa\"{\i} Telescope (CFHT) detection images obtained with {\sc megacam} \citep{Boulade03} to derive the exact trace of the supernova on the spectrogram. This trace is then used as a guide for extraction. Here, "true" photometric information of the host contribution in each pixel, obtained from CFHT Legacy Survey (CFHTLS) deep reference images in various photometric bands, is used, when possible, at the stage of the extraction. 
In the favourable cases of large enough host/SN separation, we do not use a 
spectral template to model the host galaxy contribution. 
The spatial profile of the host, measured on the reference image along columns parallel to the slit, is used instead. 

Technical choices sometimes different from the standard 
ones have been investigated to select the most efficient and 
robust ones necessary for a clean and final extraction. We have automated
the reduction pipeline as much as possible to reduce the operator work load as
well as unavoidable human errors. Besides the technical aspects of this new approach, in all the reduction and extraction processing, one of our main concerns is to control and propagate the noise level estimation from 
raw data up to the final reduced spectrum in the cleanest way possible. In particular, re-sampling of data that introduce correlations among pixels is kept to a minimum and is done as late as possible in order to preserve as long as possible the independence of pixels. This allows us to extract a clean, optimal\footnote{In fact, our extraction is not strictly optimal because of the
PSF modelling as a Gaussian (see below). However, we are close to optimality (within a few percent of minimum variance). In the following, the word 'optimal' stands for 'close to optimal'.} signal-to noise spectrum, and in about
70\% of the currently treated spectra (3rd year VLT spectra of the SNLS, see \citealt{Balland08}), this is a  host-free supernova spectrum. We discuss in this paper the improvement in terms of S/N over more standard extractions.

Identification of the supernova type is then performed by combining spectral and light curve informations using the spectrophotometric template SALT2 developed by \citet{Guy07}. In this approach, all available information on a given supernova (both photometric and spectral) is used to ease the identification.
This contrasts with the more straightforward method of a two component spectral model. As underlined in, e.g., \citet{Hook05,Lidman05,Balland06,Balland07}, a difficulty of this approach is the possible confusion of \Iae spectra past maximum with Type Ic supernovae (SNe~Ic) at an earlier phase. Using the photometric phase as a constraint is crucial to alleviate this degeneracy. 

In this paper, we present the techniques developed for the spectral extraction and the identification of our spectra. {We illustrate our results with a few sample spectra obtained at VLT, as part of two large spectroscopic programmes running from June 2003 to September 2007, that are being processed using this technique  \citep{Balland08}}. In Sections 2, 3 and 4, we describe the 
PHASE technique and test it on simulated data. In Section 5, we illustrate PHASE results and discuss the improvements over standard extractions. In Section 6, the identification method of SNLS spectra using the SALT2 spectral template is described and results are discussed on a few
examples. Discussion and conclusion are in Section 7 and Section 8, respectively.

\section{2D spectrum calibration using PHASE}

Spectral extraction usually requires human interaction
to define the locus on the frame of the object to be extracted/measured.
Spectra are often distorted along the
dispersion axis, and the extraction algorithm
must be sophisticated enough to find and follow the source trace
along this axis. Technically, one first specifies the source
location as a set of
pixel coordinates that is subsequently corrected during the extraction 
process.

Such a procedure, if relevant in the case of resolved
sources, is not wanted
for overlapping sources such as
a point-like supernova on its extended host galaxy.
Beside the fact that both the supernova and its underlying
host light are extracted together
, the position and slope of the supernova trace itself can 
not be easily retrieved.
Indeed, usual algorithms such as \citet{Horne86} seek the flux centroid,
and might end up pushing the position of the supernova towards the galaxy 
brighter core.

In order to avoid these pitfalls, we use the fact that
SNLS produces deep images of the regions in which supernovae
explode. The supernova location and magnitudes in four photometric bands ($g_M,r_M,i_M,z_M$) are known.
Consequently, given this photometric information, it is possible to 
locate the supernova on the spectrogram at sub-pixel accuracy, 
to trace its spectrum, 
and to perform an optimised extraction able to 
separately measure the supernova and host galaxy fluxes along the CCD frame.

\subsection{FORS1 long slit spectra calibration}
\label{sec:fors1lss}

During the first VLT large programme (2003-2005), SNLS spectra have been 
acquired in long slit mode (LSS) with the FORS1 instrument in service mode. 
A preliminary assessment of the spectroscopic data was performed within
a few days in order to confirm the validity of the candidate.
For this ``real-time" analysis of the spectra \citep{Basa08}, the SNLS team
used a combination of a dedicated reduction pipeline based on the ESO-MIDAS 
software, a proper \citet{Horne86} extraction package and a template 
supernova and galaxy fitting software (${\cal SN}$-fit, \citealt{Sainton04a,Balland06})  
to perform
an on-line identification and redshift measurement. 

The FORS1 detector is a 2k$\times$2k CCD. To acquire supernova spectra, we used it with the standard collimator 
and mainly with the 300V grism\footnote{And sometimes the 300I grism for the farthest supernovae, 
$z > 0.8$}, 
along with the GG435 order-sorting filter.
 The pixel scale is $0.2''$ along the spatial axis ($Y$), and $2.65\, \AA$ along the dispersion axis ($X$).
PHASE has initially been implemented to treat the
long slit spectra taken on FORS1.
The multi-object (MOS) mode is currently marginally supported (the majority of standard stars were observed in MOS mode) but its full support is in progress and will be used to extract the MOS spectra, mostly obtained during the second 
SNLS-VLT large programme (2005-2007).

In the following, we focus on the main features of the data processing.
Technical details can be found in \citet{Baumont07}.

For calibration purposes, we built bias master frames as 
median of bias frames grouped per trimester 
($\sim 50$ frames).
The scan subtracted r.m.s. of each pixel value was clipped
from $8\sigma$ deviant events, where $\sigma$ is the read-out-noise (RON) level taken from the raw frame header. With this threshold, most cosmic rays were efficiently removed. 

Normalised flat-fields were produced by grouping frames
per run (per lunation). 
After an iterative median with a $5\sigma$ clipping suitable for
removing most of bad pixels\footnote{This time, $\sigma$ also includes the photon noise},
the internal lamp spectrum was estimated in 16 spatial bands along the $Y$ axis
and used to produce normalised flat-fields that reflect sensitivity 
fluctuation patterns in those regions.

The dispersion function for science calibration was
computed using a least square technique. 
The pixel position $X_{i,j}$ of 14 isolated emission lines of Hg, He and Ar
was measured in 16 regions and averaged along the Y axis.
Those lines were selected as the brightest and least blended ones.
A set of nine dispersion coefficients (4th order in $X^i$, 2nd order
in $Y^i$ and in $XY^i$) were adjusted to the 
$X_{i,j}$ values.
This set turned out to be adequate for reproducing the dispersion solution.
The 4th order coefficient $C_{x^4}$ along X has been added 
(it is not used by the Real-Time ESO pipeline) to further reduce the residuals. 
The residual r.m.s. is typically $0.2\, \AA$ and reaches
$0.8\, \AA$ at the borders of the frame.

We chose to use a single response function per
UT (FORS1 moved from UT1 to UT2 on June 2005) by combining
all individual response functions derived for every standard 
star observation. 
With this procedure, the night to night sky transparency variation is 
not corrected for. 
We prefer to use a uniform, easy to control, average calibration for the full set, rather than calibrate an observation using the response function of a different night (ESO does not perform standard star observations for every night). As SNLS spectra are primarily used for the purpose of identification, 
we do not need absolute spectrophotometric calibration.
Second order residual flux was sometimes found for the bluest standard star spectra. In this case, we estimated it and accordingly 
corrected the response function.
The average $H_20$  and $O_2$ atmospheric absorption spectra
were computed in appropriate bands 
([5800:6000], [7100:7350], [8100:8400], [8900:9850] $\AA$ for $H_2O$, 
and [6250:6350], [6800:7000], [7550:7750] $\AA$ for $O_2$) and used for further 
correction (see Section \ref{sec:speccalib}).
As for the instrumental response, we used an average spectrum of the
absorption, computed as the average ratio 
of the observed spectrum to the reference spectrum of the standard
star, multiplied by the instrumental response interpolated 
in a given absorption band.

\subsection{Combined 2D spectrogram}

In general, we took $N_{img}$=2 to 5 exposures of 900s of our targets, with small offsets along the slit. 
We combined these $N_{img}$ frames of a given target to produce a clean, cosmic free sky subtracted 2D spectrogram to be used for extraction.

For cosmic ray removal prior to extraction, we used
a temporal median filtering. The sky spectrum was obtained in 16 spatial 
regions by an iterative median
projection along the $Y$ axis. These regions are of equal width and cover the
full extension of the CCD along the $Y$ axis. Bright sources were detected
as rows containing more than 5\% of 5$\sigma$ deviant pixels and were
avoided. The $16$ sky spectra were then averaged in wavelength space 
to yield the
sky spectrum model for sky subtraction. Note that bias subtraction and 
flat-fielding were applied simultaneously during the first projection along 
$Y$ in the iterative process.

For a given SN candidate, bias subtraction, flat fielding and sky subtraction
was then performed on each set of $N_{img}$ 
pixel values. Then a robust, seeing-weighted 
average of each calibrated set of pixels was applied to obtain one single 
output pixel value.
The seeing was computed from the active optic wavefront sensor system 
on the guiding star, recorded in the ESO logs of observation. 
In rare cases (for a few standard star observations), those guiding star 
logs are lacking and the Differential IMage Monitor (DIMM) value was used instead. This latter value is accurate within 20\%.

\begin{figure*}
\begin{center}
\resizebox{18cm}{2cm}{\includegraphics{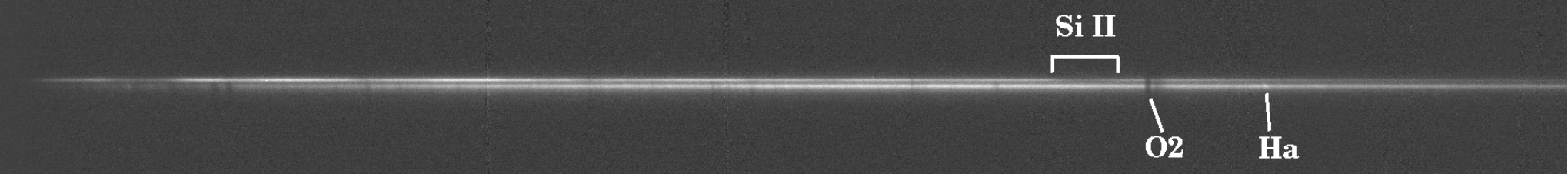}}\\[2mm]
\resizebox{18cm}{2cm}{\includegraphics{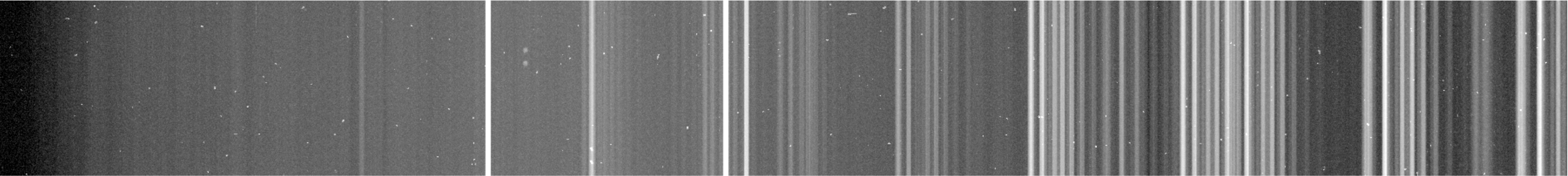}}\\[2mm]

\end{center}
\caption{PHASE 2D combined spectrograms for SN~04D1dc. Top: the SN (top) and host (bottom) spectra after reduction. The main noticeable features in the SN and host spectra are indicated. Bottom: the corresponding 2D noise map, including the sky model. Spectra range from
4200 $\AA$ (left) to 9000 $\AA$ (right). Note the atmospheric absorption features affecting both the SN and host spectra.}
\vspace*{3cm}
\label{fig:2Ds}
\end{figure*}

Figure \ref{fig:2Ds} shows the PHASE output 2D frames for SN~04D1dc\footnote{The SNLS naming scheme is the following: each supernova name is composed by the year of its detection (03-08), followed by the field name (D1-D4) and two letters in the order of discovery} taken as an illustration of the results produced by the reduction technique presented above. SN~04D1dc is a Type Ia supernova (SN~Ia) at maximum light at $z=0.211$ well separated from its host core ($d_{SN-host\  centre}=1.18''$). The top panel shows the well separated SN (top) and host (bottom) spectra. Sky subtraction residuals appear as vertical narrow bands slightly visible at the
locus of the most luminous sky emission. Atmospheric absorption features 
affecting both spectra are clearly seen at several places, such as the strong 
$O_2$ absorption at 7600 $\AA$. The emission and absorption features of the SN spectrum are 
seen as a modulation in the intensity of the top spectrum. The deep 
\ion{Si}{\sc ii} absorption at $\sim 6150$ \AA~ is visible to the blue part of the $O_2$ absorption (see labels in top panel of Fig. \ref{fig:2Ds}). In the host
spectrum, \ion{H}{$\alpha$} emission is seen as a bright spot longward of $O_2$. Confusion of an emission line with a cosmic is very unlikely 
(a cosmic hit could mimic an emission line and extend spatially but 
it should be faint enough for it not to have been rejected, i.e. $<5\sigma$ of the average sky background). However, at high redshift ($z\geq 0.8$), the host 
flux is weak and it is more difficult to tell from the sole 2D spectrum whether
a given spot is an emission line, a statistical fluctuation, or a cosmic residual. In the present case, we can check in the 2D noise map shown in the lower
panel of Fig. \ref{fig:2Ds} that the observed bright spot in the spectrum is not a cosmic subtraction residual. Indeed, if a cosmic hit the CCD at this location, it would be present in this 2D noise map, which is not the case. This illustrates how inspection of the 2D noise map helps at discriminating true host lines from cosmic hits. 

\section{A model for PHASE extraction}

The basic idea for the extraction from the clean 2D spectrogram
is to define beforehand the distribution of the sources that we expect
to find in the frame and to recover their exact position (or trace) on 
the spectrogram. To achieve this, we rely on the comparison between the source spatial profiles obtained from deep CFHTLS reference images in three {\sc megacam} photometric bands ($g_M$, $r_M$ and $i_M$ with Grism 300V\footnote{And $r_M$, $i_M$ and $z_M$ with Grism 300I.}) along the slit direction (the photometric profiles), with the corresponding profiles obtained from the VLT Long Slit spectrograms (the spectroscopic profiles). Then, given the expected spatial profile of the sources, 
a least square linear inversion of the pixel counts on each column, 
corresponding  to a given $\lambda$, gives the flux of every component at 
this wavelength. In the following, we describe how we implement this procedure
in PHASE.

\subsection{Photometric priors}
\label{sec:photopriors}

To built the photometric profiles, we project
the region of the sky covered by the slit at its position angle 
from the deep reference images in all of the {\sc megacam} photometric
filters covered by the FORS1 order-sorting filter
used (i.e, g$_{M}$, r$_{M}$ and i$_{M}$ for GG435 used with the 300V grism).
This is done using SWarp \citep{Bertin07}. This yields three 1D spatial 
profiles that correspond to the expected {\em galactic} profiles in each 
filter. 

The supernova coordinates, accurately measured when building its light curves,
are used to place the supernova component on these profiles.
Given that supernovae are point-like sources, the supernova component
is modelled as a Gaussian profile whose width is the average 
seeing of the spectroscopic observation. 
As the supernova flux is not accounted for in the CFHTLS deep reference 
images, we add it ``by hand'' to the photometric profiles.  
To do this, the flux in each band is interpolated from the 
supernova light curves at the time of spectroscopy. 

\begin{figure*}
\begin{center}
\resizebox{17cm}{10cm}{\includegraphics{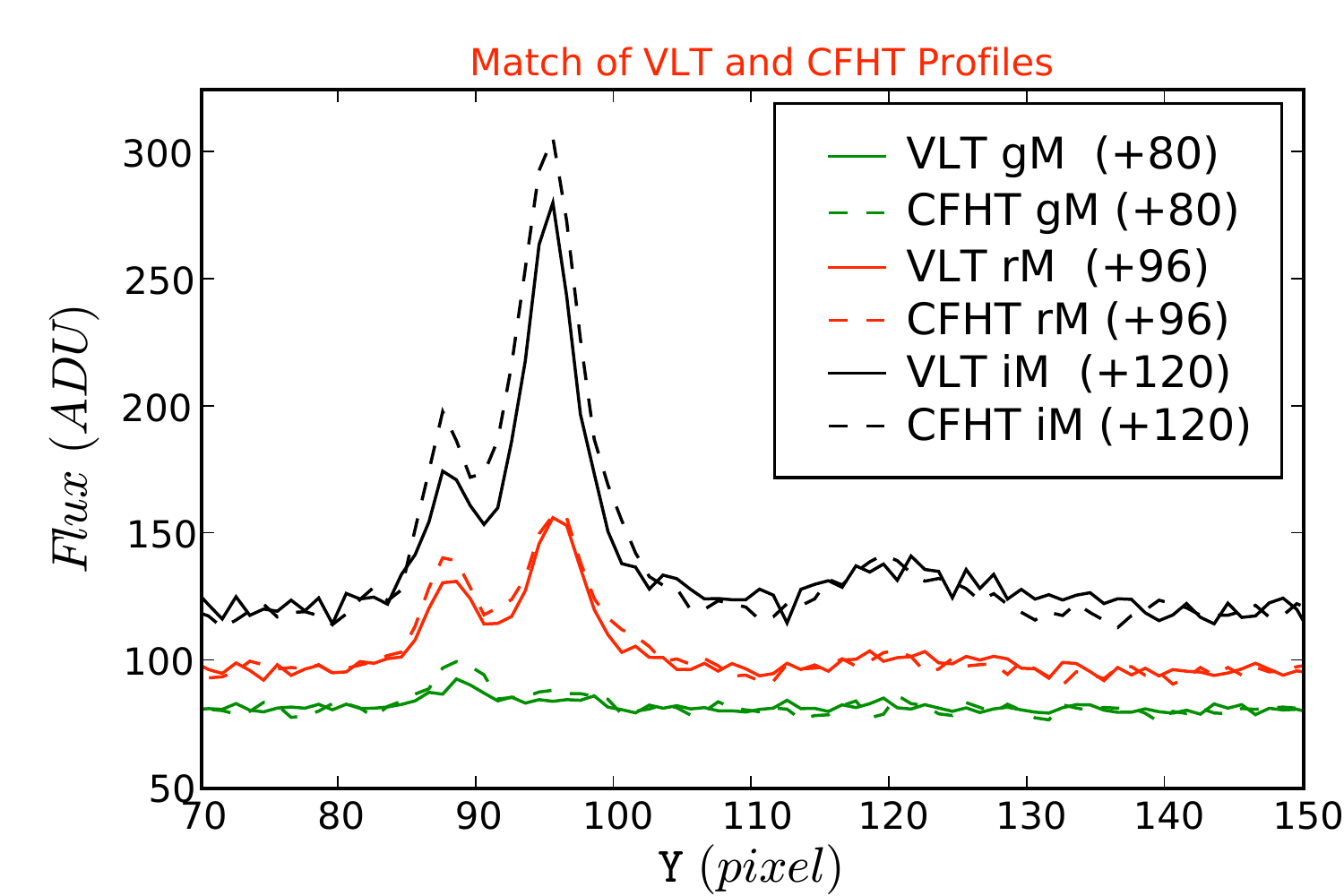}}\\[2mm]
\end{center}
\caption{Match of the VLT (solid lines) and the CFHT (dashed lines) spatial profiles of SN 04D4it in
three photometric bands (from bottom to top: $g_M$, $r_M$, $i_M$) arbitrarily shifted with respect to zero flux for the sake of clarity. The VLT profiles
are obtained by averaging the spectrum in each photometric band. The CFHT profiles are measured on deep reference images. The correlation function of
the two profiles is computed for each band and the spatial shifts are
derived at the maximum of the correlation functions.}
\vfill
\label{fig:SpectroPhotoMatch}
\end{figure*}

The photometric profiles are then compared to the spatial profiles
averaged on the 2D VLT spectrogram in wavelength ranges corresponding to
the photometric bands.
For each band, the spatial shift 
between the two derived profiles (photometric and
spectroscopic) is then computed at the maximum of their 
correlation function.
Figure \ref{fig:SpectroPhotoMatch} illustrates the result of matching
the VLT (solid lines) and the CFHT (dashed lines) profiles in $g_M$, $r_M$ and $i_M$ (from bottom to top) for SN 04D4it, chosen as a test case.  
Correlating the two profiles yields one shift for each photometric band 
whose mean $\lambda$ corresponds to a given column $X_{Band}$ of the spectrogram. Performing a {\em linear} fit of these shifts as a function of $X$ yields
two parameters\footnote{Note it would be possible to compute the trace curvature as a second 
order term in the fitting function, but the deviation from linearity is only a 
fraction ($\sim 0.1$ pix) of a pixel. We prefer to over-constrain the fit since a given band 
(e.g., g$_{M}$ at high redshift) might contain too little flux to get 
an accurate shift value.}:
\begin{itemize}
\item 1) the spatial shift between the supernova location and the central
row of the spectrogram, at the central column, and 
\item 2) the slope of the supernova location along the dispersion axis $X$.
\end{itemize}

The maximum tolerated spatial shift is 20 pixels.
In average, the slope is of
 $10^{-3}$ pixel/pixel,
corresponding to a 2 pixel drift from border to border of the frame.

This procedure hence allows us to measure precisely (at sub-pixel accuracy)
the position and slope of the sources spectral trace on the 2D spectrogram.
We now use this information to perform the extraction.

\subsection{Building a multi-component model}

For each column flux $F_i(\Yp)=F(\Xp_i,\Yp)$ of the 2D spectrogram, 
we build a $N$-component spatial model as the sum of the profiles 
$\mathcal{C}_k(\Yp)$ of individual objects in the slit, shifted by 
a quantity $\delta\Yp_i$ according to the previously derived trace 
equation. Their total flux is normalised to unity.  
A flux $f_k$ is assigned to the $k^{th}$ component of the model and  
a uniform ``background" component is added to account for possible sky subtraction residuals. The model for each column $i$ reads:
\begin{equation}
\label{eq1}
{\mathcal M}_i(\Yp)=\sum_k f_k {\mathcal C}_k(\Yp-\delta\Yp_i).
\end{equation}

The flux $f_k$ of each component is estimated by a $\chi^2$ minimisation where:
\begin{equation}
\label{eq2}
\chi^2=\sum_j\Big(\frac{F_i(\Yp_j)-{\mathcal M}_i(\Yp_j)}{\sigma_j}\Big)^2.
\end{equation}
Here, $\sigma_j$ is the error associated with the column pixel data.
The minimisation condition is equivalent to solving the following equation for each column:

\[ \left[ \begin{array}{ccc}
    \sum_j \frac{\mathcal{C}_0(\Yp_j')^2}{\sigma_j^2} & \cdots & \sum_j \frac{\mathcal{C}_0(\Yp_j')\mathcal{C}_N(\Yp_j')}{\sigma_j^2} \\
    \vdots & \ddots & \vdots \\
    \sum_j \frac{\mathcal{C}_N(\Yp_j')\mathcal{C}_0(\Yp_j')}{\sigma_j^2} & \cdots & \sum_j \frac{\mathcal{C}_N(\Yp_j')^2}{\sigma_j^2} 
  \end{array} \right] 
  \left[ \begin{array}{c} f_0 \\ \vdots \\ f_N \end{array} \right] = \]
\[\hspace*{5cm}  \left[ \begin{array}{c}
    \sum_j \frac{\mathcal{C}_0(\Yp_j')F_i(\Yp_j)}{\sigma_j^2} \\
    \vdots \\
    \sum_j \frac{\mathcal{C}_N(\Yp_j')F_i(\Yp_j)}{\sigma_j^2} 
  \end{array} \right],
\]

where $\Yp_j' = \Yp_j-\delta\Yp_i$ corresponds to the profiles shifted according to the trace inclination.

Much care must be taken to model the source profiles, since their
accuracy will determine the quality of the extracted spectra.
In particular, overlapping sources is a source of anti-correlated noise
among components.

\subsubsection{Setting the galaxy profiles}
\label{sec:GalPhaseModel}

If the extraction window is narrow enough 
to ensure that no field galaxy other than the host is present, 
the galaxy profile and the supernova Gaussian can be 
used to perform an extraction with only {\em two} components (plus a uniform
background component).

However, using a wide extraction window is preferable for 
a better sky residual estimation. This often implies that other
object spectra are present in the window. 
Even in the case of a narrow extraction window, close galaxy pairs 
are sometimes present. In all these cases, we need to define
a multi (greater than 2) component model. 
Moreover, for resolved spiral galaxies, the spiral arm
spectrum is not the same as the core spectrum.
In particular, nebular emission lines are much stronger
in the arms due to the enhanced star burst activity. The \ion{H}{\sc ii} regions, ionised by the strong UV radiation from young and massive stars, produce
these nebular emission lines \citep{Osterbrok89}. 
Improper modeling of the arm and
core structure as a single component leads PHASE to fail to assign 
the correct flux to the SN and to the galaxy. At the wavelengths of
these nebular emissions, the host spatial profile does not correspond
to the host profile averaged over the whole spectral range. In the
case of an SN located in one arm, using the average profile
to model the host would lead to underestimate the host flux at these
wavelengths, and consequently, to overestimate the supernova flux.
In such cases, it is necessary to model the arms and the core as two 
different components.
Considering that high-redshift host galaxies are 
usually unresolved above a redshift of 0.7, we 
distinguish 3 types of galaxy profiles:
\begin{itemize}
  \item $PSF$ : unresolved, point-like galaxies.
  \item $EXT$ : extended, regularly shaped profiles, e.g. ellipticals.
  \item $Mix$ : irregularly shaped profiles, e.g. galaxies with spiral arms. In practise, this profile is modelled as a PSF plus an EXT profiles.
\end{itemize}

In PHASE, cuts on flux, galactic compactness, extension minimum level
and colour variation between the central Gaussian core
(at seeing width) and the possible extension (e.g., arms)
can be adjusted to favour one type over the others and thus adapt to the 
specific extraction case encountered.
Nevertheless, default cuts are set to permit an automatic
treatment of most spectra.  Their values are chosen to yield 
the host type model (Mix, EXT or PSF) giving the best extraction (avoiding host contamination and degeneracy between the SN and host components)
in most cases. 
For extended sources, we use the ``bolometric" profile 
(that is the sum of the galactic profiles in all covered filters)
as a model.
In order to use a noiseless extraction profile, it is filtered with
a Gaussian of width half the seeing, oversampled
at a tenth of a pixel to ease the subsequent shifting
of the profile to follow the trace slope.
Note that using the projected 1D profile is a source of
confusion when close pairs of galaxies happen to
connect once projected along the given polar angle.
However, sources are well
defined for the subsequent extraction in $\sim$95\% of the cases\footnote{{We refer here to the whole set of VLT SNe~Ia spectra extracted using PHASE and presented in \cite{Balland08}}}, and this is
not a major concern.

\subsubsection{Modelling the SN component}
\label{sec:ModelSN}

As mentioned in Section \ref{sec:photopriors}, the candidate supernova is an 
additional point-like source at the estimated position and of width equal to 
the seeing. The seeing variation with wavelength was estimated from standard star
observations, as a power law of index $-0.3$. This corresponds to a $\sim 20$ \% variation of the
Full Width at Half Maximum (FWHM) from 4000 \AA~ to 9200 \AA.
Note that our power law value differs from $-0.2$, the value expected
when assuming Kolmogorov turbulence \citep{Schroeder87} but is in agreement with the value
found by \citet{Blondin05} on FORS1 spectra. 
Note that we use this law for a distant PSF host as well. However, applying the same correction to 
extended sources would require a reference image of better seeing, and would 
only have a minor effect on the profile, so we do not perform it.

To avoid degeneracy between two overlapping point-like sources 
(e.g., the supernova and an unresolved host galaxy),
 a cut of a fifth of the seeing is imposed on the separation of the two
point-like sources. 
For separation values lower than this cut, the SN component is merged to 
the unresolved galaxy and no separate extraction is possible.
As expected, this situation 
occurs with increasing frequency at higher redshifts and only the sum of the supernova and of the host spectra can then be obtained.
This situation occurs in about 30\% of cases (see \citealt{Balland08}) and is further
referred to as the ``{\em SNGAL}" case.
However, the faintest hosts are not identified as sources
since their level does not reach the detection threshold fixed in PHASE,
and their spectra will only slightly contaminate the SN signal. When such a case occurs (or when a clear separation between the SN and the galaxy exists), we refer to it as a ``{\em SN}" case.

\section{Testing the validity of PHASE extractions}
\label{sec:simul}

\begin{figure*}
\begin{center}
\resizebox{10cm}{8cm}{\includegraphics{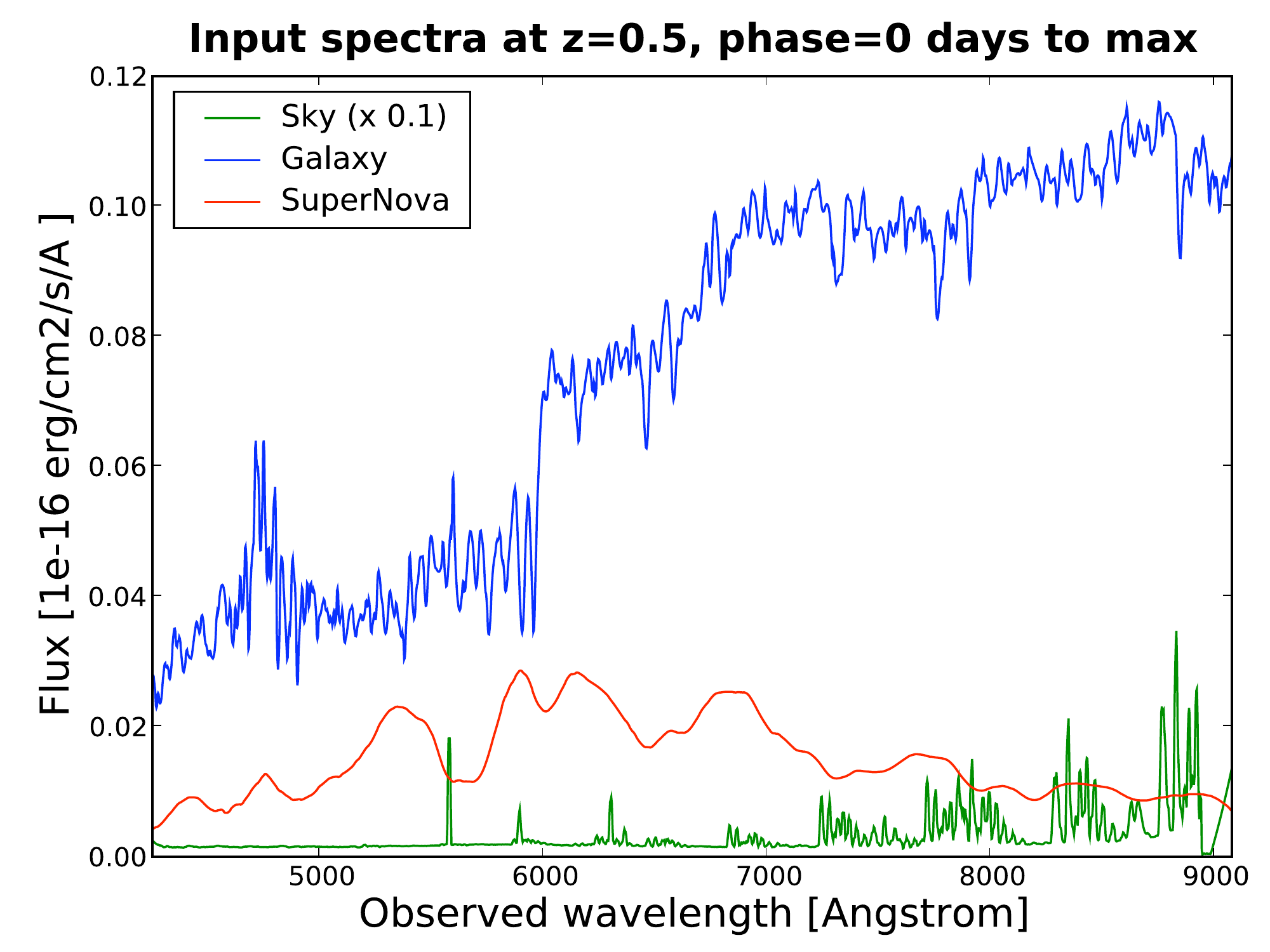}}\\[2mm]
\end{center}
\caption{The host (blue) and sky (green) spectra used in the simulations testing the
validity of PHASE extractions. The host is a Sb type from \citet{Kinney96} and
the sky background is taken from a real FORS1 exposure and is used to model the noise of the 2D input spectra. The flux of the Sb template is scaled to the one
of NGC~6181 \citep{Kennicutt92} at a distance of 34 Mpcs and its spatial profile on the
synthetic 2D spectrogram is fixed from the angular size of NGC~6181 using the
surface brightness profile of \citet{Ratnatunga99}. The supernova spectrum is modelled from the templates of \citet{Nobili03}, with phases ranging from -10 to 10 days. Here, the supernova template at maximum light is shown (red line) as an example.}

\vfill
\label{fig:inputspec}
\end{figure*}

In order to assess the validity of PHASE extractions, we have
performed
a series of simulations to test the impact of various parameters
(supernova/host separation, phase and redshift of the supernova, seeing)
on the recovered flux for each component.
We start from \citet{Nobili03} SNe~Ia template spectra with phases ranging
from -10 to +10 days relative to maximum light. We assume that the supernova
host is resolved and use the \citet{Kinney96} Sb template to model its 
spectral energy distribution.
We follow \citet{Blondin05} and generate the input synthetic 2D spectra 
(combined and sky background subtracted, for 3 exposures of 750 seconds) 
within the FORS1 instrumental environment. 
The profile of the host galaxy on the 2D synthetic spectrogram is fixed from the 
angular size of NGC 6181 \citep{Kennicutt92}
and using, as in \citet{Blondin05}, the \citet{Ratnatunga99} surface brightness profile. 
The flux of the host template is scaled to the one of NGC 6181 
at a distance of 34 Mpcs. A model
for the sky background (taken from a real FORS1 exposure) is 
used to model the noise of the input 2D spectra 
(in addition to the RON and to the shot noise of the sources). 
A curvature of the spectral trace (not accounted for during PHASE extraction)
is modelled, introducing a 0.1 pixel offset from centre to border of the
dispersion axis, consistent with the FORS1 curvature measured on standard star
exposures.
Figure \ref{fig:inputspec} shows the input host and sky spectra, as well
as the supernova spectrum at maximum light, for $z=0.5$.

We extract the signal with PHASE, imposing the photometric model.
We perform 3 sets of simulations varying either the redshift ($0.2<z<1$), the SN phase ($-10<\phi<+10$ days) or the seeing ($0.4''<FWHM<1.4''$). In
each case, the host/SN separation $d_{SN/Host\ centre}$ varies between 
0 and 2$''$. 
The non-varying parameters are set to the default values $z^0=0.5$,
$d_{SN/Host\ centre}^0=1''$ and $FWHM^0=0.75''$.

\begin{figure*}
\begin{center}
\resizebox{10cm}{8cm}{\includegraphics{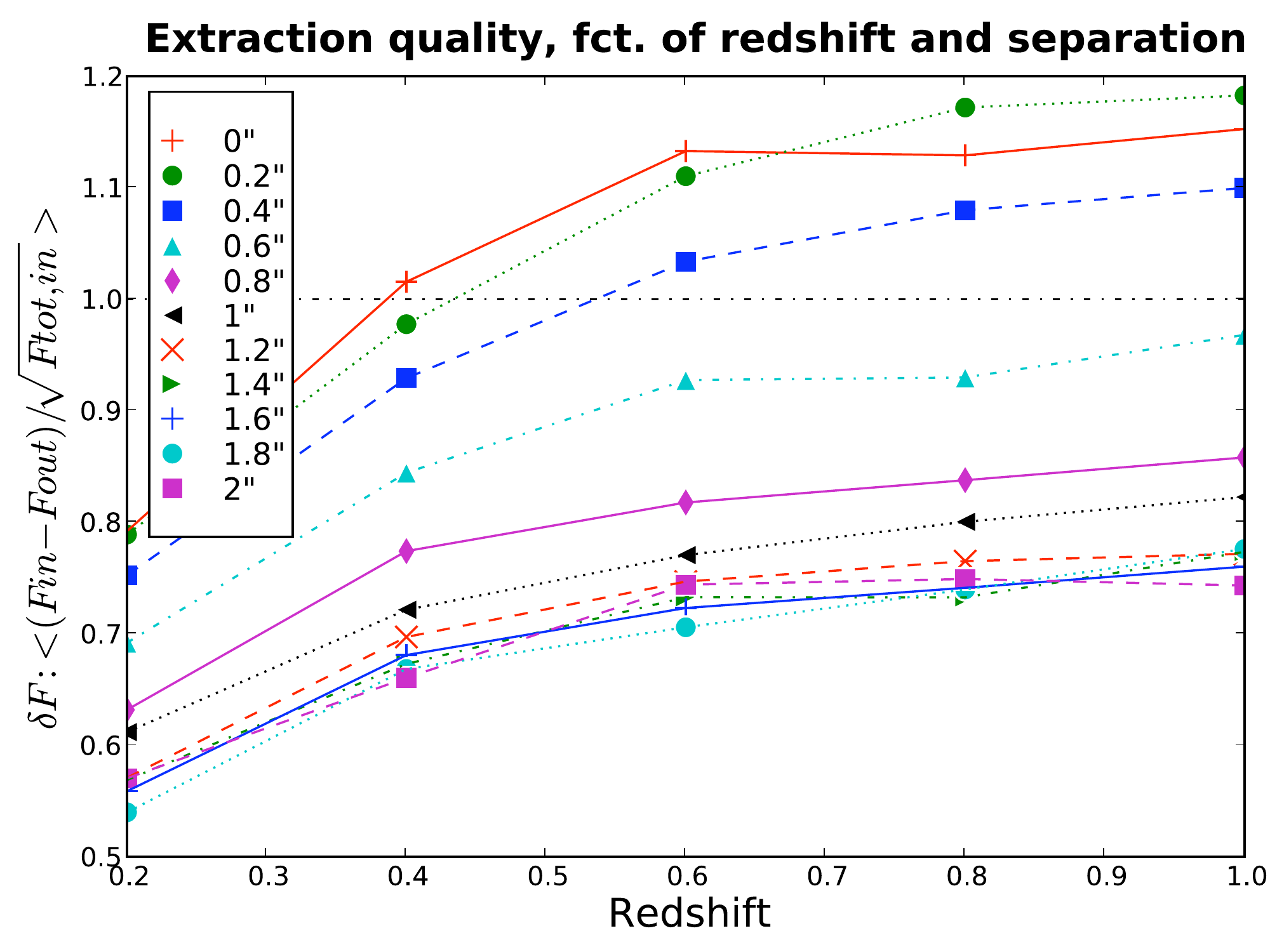}}\\[2mm]
\end{center}
\caption{Residual $\delta F$ of a PHASE extraction performed on simulated data as a function
of redshift. Values of $\delta F$ lower than 1 indicate that the flux has been recovered
to the statistical noise limit. Symbols show $\delta F$ values for SN/host separations ranging from 0 to 2$''$. For $z>0.5$ and  $d_{SN/Host\ centre}<0.4''$, $\delta F$ is greater than 1 as 
the galaxy becomes
unresolved so that, for low separation, the degeneracy between the two
component profiles is high. In practise, in such cases, PHASE would extract the
two components together and not separately. }

\vfill
\label{fig:simul1}
\end{figure*}

\begin{figure*}
\begin{center}
\resizebox{10cm}{8cm}{\includegraphics{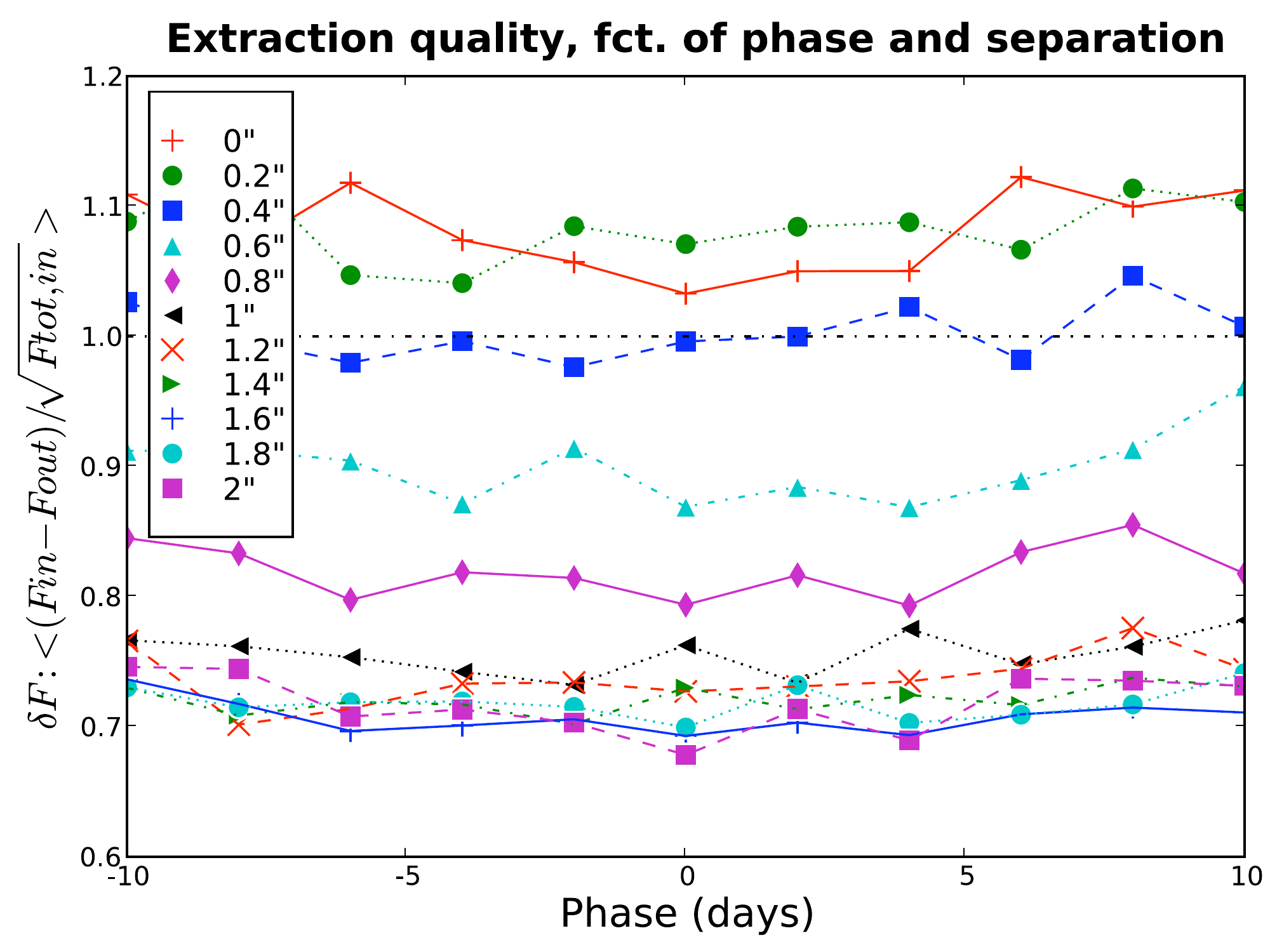}}\\[2mm]
\end{center}
\caption{Same as Fig. \ref{fig:simul1} as a function of the phase of the simulated supernova}

\vfill
\label{fig:simul2}
\end{figure*}

\begin{figure*}
\begin{center}
\resizebox{10cm}{8cm}{\includegraphics{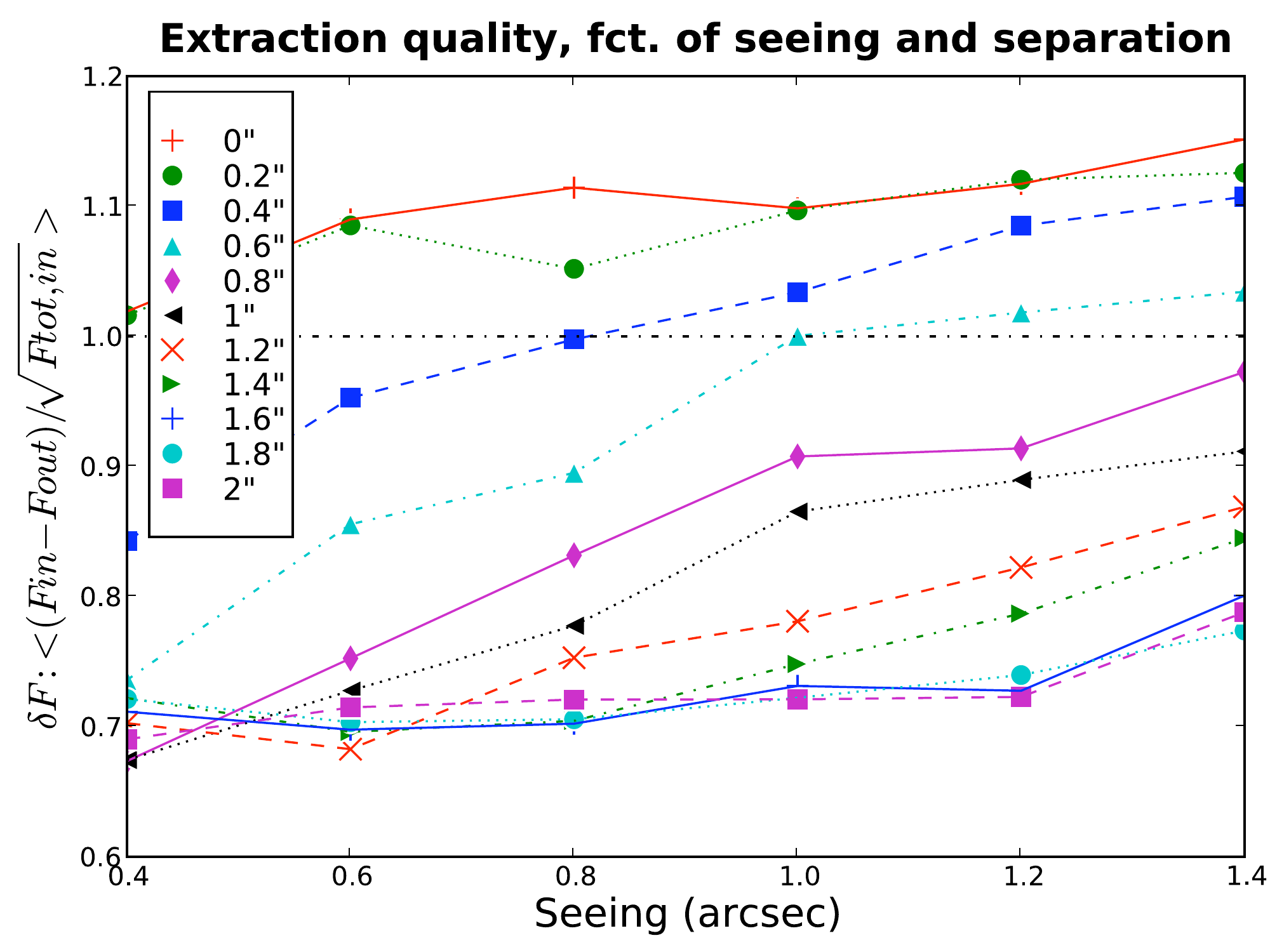}}\\[2mm]
\end{center}
\caption{Same as Fig. \ref{fig:simul1} as a function of the seeing of the simulated VLT spectrum.}

\vfill
\label{fig:simul3}
\end{figure*}

To evaluate the quality of the flux recovery, we adopt the criterion
of \citet{Blondin05} based on the residual $\delta F=
|F_{out}-F_{in}|/\sqrt(F_{tot,in})$, where $F_{out}$ is the recovered supernova flux, $F_{in}$ the input supernova flux, and $F_{tot,in}$ takes into account the supernova+underlying host+sky background. If $\delta F<1$,
the input signal has been restored to the statistical noise limit.
Figures \ref{fig:simul1} to \ref{fig:simul3} present $\delta F$
as a function of redshift, phase and seeing.
We see in Fig. \ref{fig:simul1} and \ref{fig:simul2} that PHASE recovers the correct flux 
($\delta F<1$) for supernova-host
separation $d_{SN/Host\ centre}$ greater than 0.5", even for high redshift ($z>0.9$) and for early or late phases.  For $z>0.5$ and  $d_{SN/Host\ centre}<0.4''$, the galaxy becomes
unresolved, so that for low separation the degeneracy between the two
component profiles is high. In fact, in such cases, PHASE would extract the
two components together (``{\em SNGAL}" case, see Section \ref{sec:ModelSN}). 

Note that the effect of the SN phase is very marginal (Fig. \ref{fig:simul2}) , indicating that low
contrast of the SN relative to the host is not critical to recover the
SN spectrum.
From inspection of the residuals, we find that the unaccounted curvature 
is the main source of contamination of the supernova spectrum by the host
spectrum, which will scale with this contrast.

From Fig. \ref{fig:simul3}, we see that for $d_{SN/Host\ centre}$ under 0.5$''$ and a seeing larger than 0.9$''$, $\delta F$ is above 1, but remains under 1.15\,.

We conclude that in a wide range of redshifts, seeing and phases, PHASE efficiently recovers the fluxes of the SNLS VLT supernova
spectra. At $z=0.5$ and for our median seeing of 0.7$''$, PHASE is able to
extract separately the supernova from its host provided their separation
is greater than $\sim 0.2''$ (1 pixel).

\section{Results}

\subsection{Examples of PHASE extractions}
\label{sec:speccalib}

\begin{figure*}[htbp]
\begin{center}

\begin{tabular}[b]{cc}
  \hspace{-5mm}
  \includegraphics[width=8.5cm]{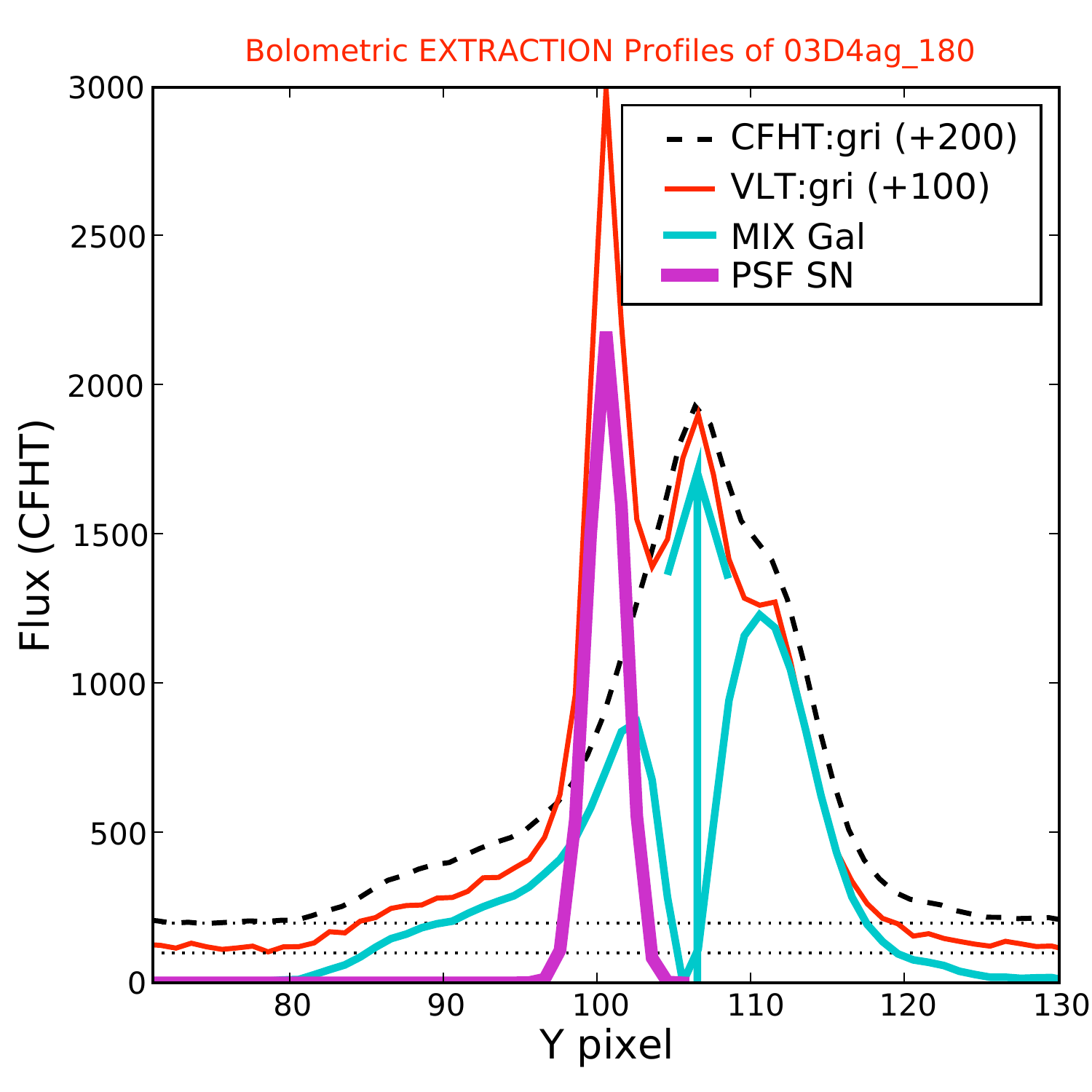}
  &
  \vspace*{-3mm}\includegraphics[width=8.8cm]{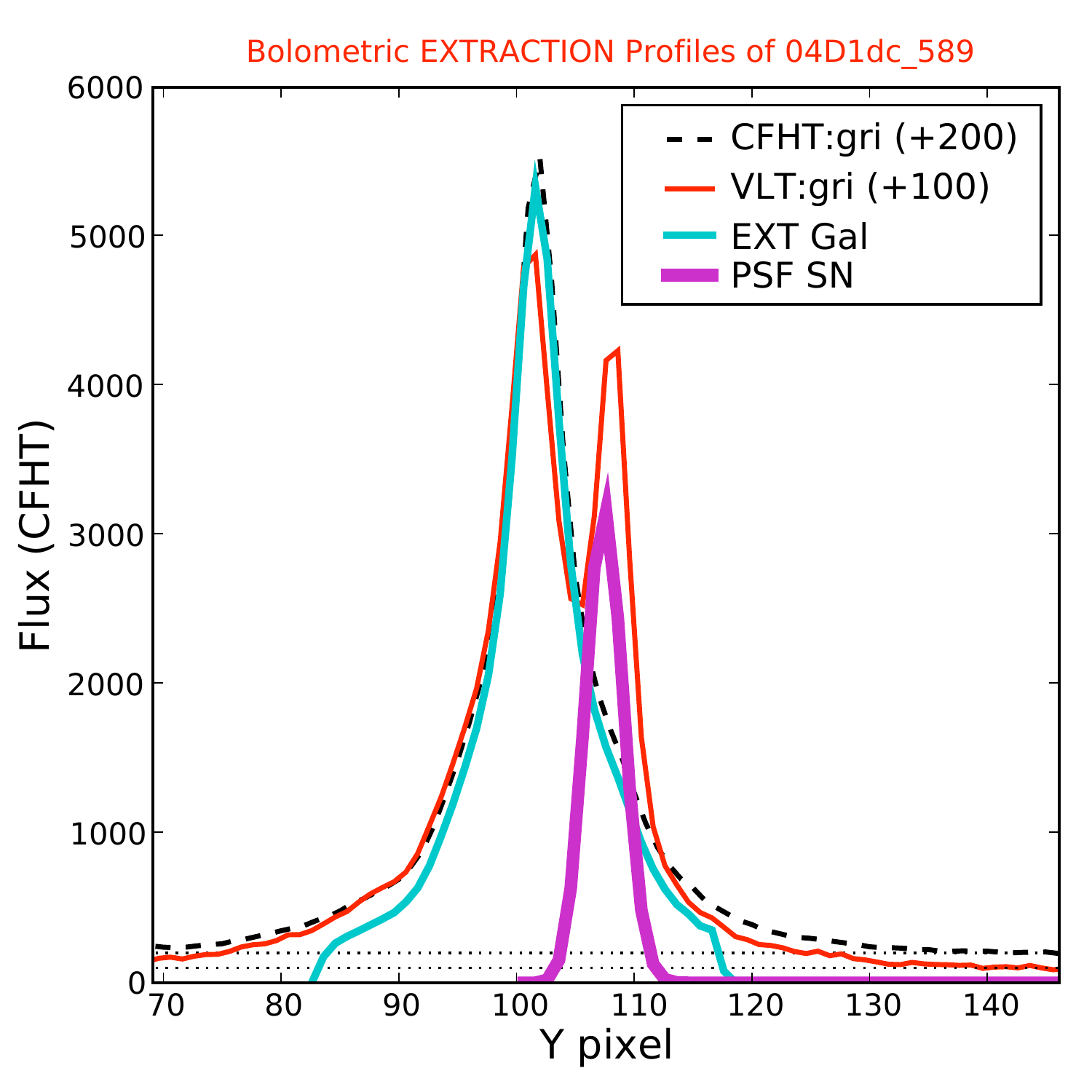}\\
\end{tabular}
\caption{PHASE extraction ``bolometric" profiles for SN~03D4ag (left) and SN~04D1dc (right). The case of SN~03D4ag illustrates a Mix host model composed of a Gaussian PSF core (represented as a blue arrow) and extended arms. The case of SN~04D1dc illustrates an EXT model. In both cases, the SN PSF model is represented as a magenta Gaussian. The horizontal dotted lines indicate the
background level arbitrarily shifted for the sake of clarity.
\label{fig:PhaseModel}
}
\end{center}
\end{figure*}

In Fig. \ref{fig:PhaseModel}, we present two examples of PHASE models. 

The left panel shows the ``bolometric" profiles from the CFHT deep reference images (black dashed curve) and from the VLT LSS spectrum (red solid curve) of SN~03D4ag, an SN~Ia at $z=0.285$ spectroscopically observed  $\approx 9$ days before
maximum light. The host is modelled as a Mix model (see Section \ref{sec:GalPhaseModel}) divided into a Gaussian core (represented by a blue arrow) and an extended asymmetric arm profile (blue solid line). The supernova is represented as a magenta Gaussian. 
We thus use a 3-component model (plus a uniform background component) to 
extract the spectrum of SN~03D4ag.

The right panel of Fig. \ref{fig:PhaseModel} shows the same ``bolometric" model in the simpler case of SN~04D1dc. Here, the host is an extended early spiral galaxy with no fine structure, in contrast to the host of SN~03D4ag. It is simply modelled by the profile measured on the CFHT deep reference image. 

In both cases, a separate extraction of the supernova and the host is possible as their separation is above the default cut of 1 pixel ( $d_{SN/Host\ centre}$=1.09$''$ for SN~03D4ag and 1.18$''$ for SN~04D1dc).

After the spectra are extracted, the flux is corrected for exposure time,
instrument response, atmospheric extinction -- including absorption from molecular oxygen and water vapour.
We also correct for slit losses under the assumption that the 
supernova
is centred in the slit and that the PSF along the dispersion axis
has the same properties as along the spatial axis (a Gaussian of FWHM equal
to the effective seeing, with a wavelength variation following a power law of
index -0.3). Note that no interpolation of contiguous pixel values
is made, neither to filter cosmic hits, 
nor to re-sample the frames at uniform wavelength bins. 
This allows an accurate estimation of the noise level, 
and does not add any further correlation among pixels.
One consequence of this procedure is that the extracted spectra have
non-uniform wavelength bins.

The top panel of Fig. \ref{fig:PSFspec} presents the extracted spectrum of SN~04D1dc (red line) along with its host galaxy (blue line). Spectra have been rebinned for visual convenience. The host spectrum exhibits the features of a typical early spiral (Sa/Sb) with [\ion{O}{\sc ii}], \ion{Ca}{H\&K}, \ion{H}{$\alpha$}, \ion{N}{\sc ii} and \ion{S}{\sc ii}. Atmospheric absorption subtraction residuals are clearly seen, both in the host and the supernova spectra around 7600 $\AA$.

\subsection{PHASE performances and limitations}
\label{sec:PhaseLimits}

In order to estimate its performances, we have applied PHASE to the extraction of all spectra of the ESO-VLT
1st large programme of the SNLS. 
The full results for the confirmed SNe~Ia will be presented in \citet{Balland08}. A total of 829 LSS science and 2169 
(896 MOS and 1273 LSS) calibration exposures have been processed to produce 
208 calibrated spectrograms of SN candidates. 
In about 90\% of cases, 
the default cuts yield an optimal extraction of the signal (including cases for which a separate extraction of the SN and host components was not possible), improved over 
standard extractions using classical techniques.
The remaining 10\% of cases are events for which
extraction cuts are not suitable and thus the component definition is not adequate.
These cases are tagged after inspection of 
the output spectrum and the residual spectrogram obtained in a first automatic extraction
with default cuts. We treat these events 
on a case-by-case basis, adjusting the cuts to the specificity of the 
situation encountered. 

PHASE performs the extraction of all the objects present in the
slit. The supernova spectrum is extracted separately from the host whenever possible (about 70\% of cases). Comparison to standard extractions (see Section \ref{sec:CompRealTime})  illustrates the efficiency of PHASE extraction within the 
limits of validity of the assumptions made. These latter include:
\begin{itemize}
\item that the coordinates of the supernova are accurate.
\item that no flux from the supernova is present in the 
reference images. 
\item that the PSF is a Gaussian of width equal to the seeing.
\item that the reference image seeing is equal or close enough to the 
combined spectrogram seeing. 
\end{itemize}

Any deviation from these assumptions results in 
weak flux losses, higher noise and supernova spectrum 
contamination by its host. In particular, variation of the PSF with 
wavelength (other than the seeing dependence that we take into account) 
might be a concern. We have checked that our simple modelling of
the PSF as a Gaussian profile yields an extraction within only a 
few percent of minimum variance. Thus, PHASE extractions are
close to optimal.

In PHASE, the profile shape and centro\"id must match as closely as possible
those of the spectrogram. 
This means that one needs a deep reference image of seeing similar 
to the spectroscopic observation. However,
the average seeing of the deep reference images is $1''$, 
while spectroscopic
observations at Paranal have an average seeing of 0.8$''$.
In this case, the exact shape of the source is not well reproduced by
the Gaussian model (the galaxy is more peaked in the spectrogram than on the reference image). Position and profile inaccuracies are reflected in
the residual spectrogram obtained when the extracted 
fluxes of all components are subtracted from the combined 
spectrogram. If the seeing of the reference image is higher than the
one of the VLT spectrogram, the residuals are positive at the 
centre of the spectrogram and negative on the edges. 
Nevertheless, for a usual $0.1$ pixel position accuracy
and a $0.1''$ seeing accuracy (of average $0.8''$),
the flux loss amounts to $5\%$, with a noise increase
of about $25\%$.

\begin{figure*}
\begin{center}
\resizebox{17cm}{10cm}{\includegraphics{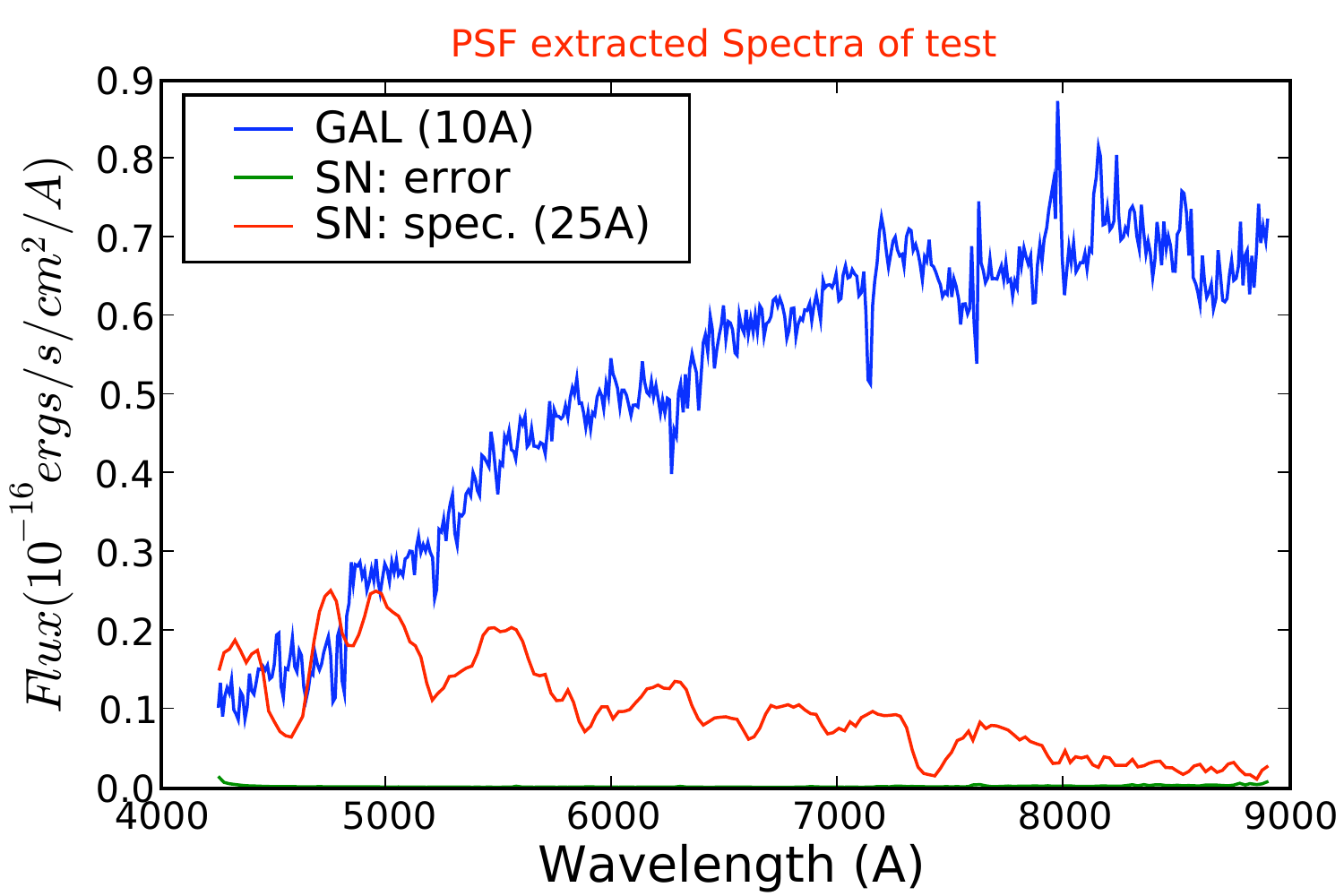}}\\[2mm]
\resizebox{18cm}{2cm}{\includegraphics{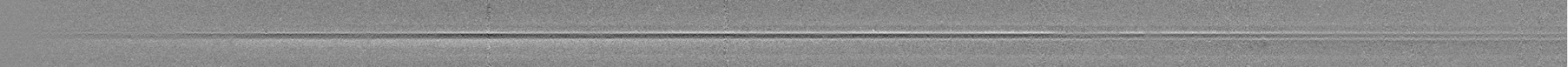}}\\[2mm]
\end{center}
\caption{Top: PSF extracted spectra for SN~04D1dc. The host spectrum (blue line) has been rebinned with 10 $\AA$ while the SN (red line) spectrum is rebinned with 25 $\AA$. Spectra have been corrected for atmospheric absorptions and are
presented in the observer frame. Bottom: extraction residual spectrogram. Position and profile inaccuracies are reflected in the residual spectrogram. Here, the supernova in the model is located $\sim 1$ pixel too close to the host centre. As a result, the model profile is not accurate: too much light is extracted at the centre and not enough on the edges. This yields negative residuals at the centre and positive residuals on the edges.}
\vfill
\label{fig:PSFspec}
\end{figure*}

The bottom panel of Fig. \ref{fig:PSFspec} illustrates
a situation for which the supernova coordinates are not accurate. Here, the
Gaussian model for the supernova is located $\sim 1$ pixel too close to its host centre. The matching of the photometric profile to
the spectroscopic profile yields too much light in the host model and
not enough in the supernova, as can be seen in the right panel of Fig. \ref{fig:PhaseModel}. The extraction yields negative residuals at the centre and positive residuals on the edges.

\begin{figure*}
\begin{center}
\resizebox{14cm}{!}{\includegraphics{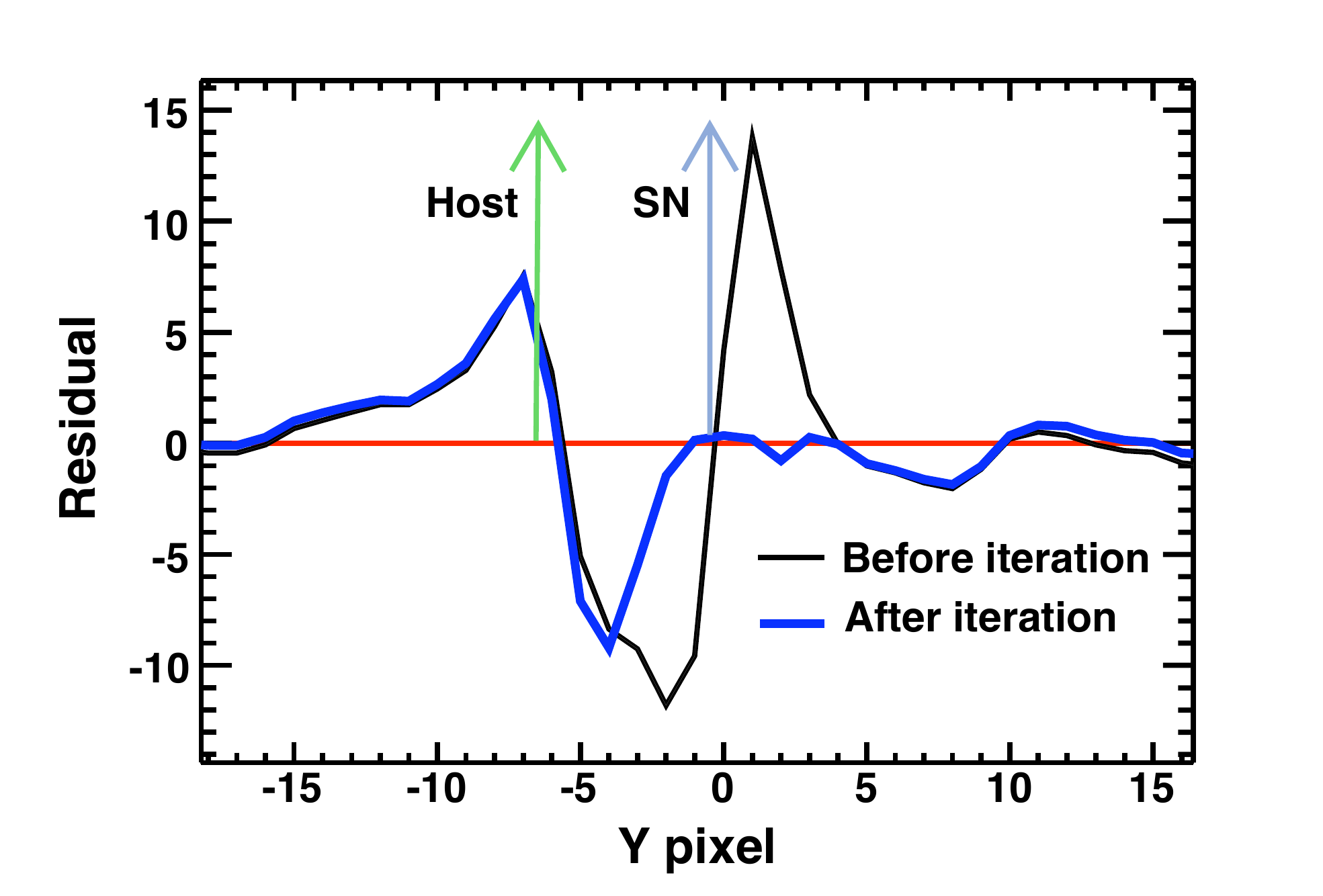}}\\[2mm]
\end{center}
\caption{Average spatial profile of the residual spectrum before (black line) and
after (thick blue line) PHASE iteration. Pixel 0 in the abscissa corresponds to the
centre of the slit. Residual scale is in ADUs/pixel. The SN and host locations along
the profile are indicated as vertical arrows. Only the position of the supernova component is allowed to vary during the iteration process.}
\label{fig:IterResidual}
\vspace*{3cm}
\end{figure*}

In order to circumvent these limitations, 
the residuals can be used to iteratively
adjust the position, slope and width of the
SN component. 
The other (galactic) components can either stay fixed, 
or move together with the SN.
Slope refinement applies to all sources and width refinement to
point-like sources only. 
Iterating in this way yields an optimal extraction 
in a few passes.
Figure \ref{fig:IterResidual} shows the improvement obtained
after iterating in this way, in the case of SN~04D1dc. The black thin
line is the average profile of the residual shown in the bottom panel
of Fig. \ref{fig:PSFspec}. The thick blue line is the average
profile of the residuals after five iterations. The host position
is kept fixed. At the end of the iterative process, the supernova
position has moved by about one pixel and its width has decreased by
$\sim 20$ \%. Inspection of Fig. \ref{fig:IterResidual} shows that
the residual flux of the supernova has disappeared after iteration. Although useful, we
do not systematically iterate, for the sake of preserving an automatic extraction.

To improve further
the residuals, one would need to refine the trace (i.e., use higher order
polynomial coefficients) and PSF models (e.g., use Moffat functions for a better fit of the effects of seeing caused by atmospheric turbulence because of their longer tails). 

Finally, regarding computer time costs, processing spectra with PHASE is 
rather efficient.
Creating the master flat-field frame is the most time expensive task of 
the calibration stage. It takes about 5 minutes of CPU time on an Intel Celeron processor running CERN-Linux to perform an iterative median on 30 frames. The definition of photometric priors and the extraction stage in itself takes only a few seconds.

\subsection{Comparison to standard extractions}
\label{sec:CompRealTime}

In this Section, we compare PHASE and standard extractions on
a few examples. Standard extractions have been performed using the 
\citet{Horne86} algorithm, in a way similar to what is done during 
the real-time processing of the SNLS-VLT spectra \citep{Basa08}.
With this method, the host and the supernova signals are often 
extracted together  
and the SN/host separation is done later when identifying the spectrum by
adjusting a model built from template spectra. This model combines a fraction of supernova and a fraction of galaxy. From a purely practical
point of view, using PHASE  
alleviates the user of a number of tasks such as the visual measure of the 
spatial shifts of the frames,
the definition of the sky estimation regions, 
and the supernova and potential host extraction regions.

\subsubsection{SN~03D4dy: a supernova resolved from its host}

\begin{figure*}
\begin{center}
\resizebox{14cm}{!}{\includegraphics{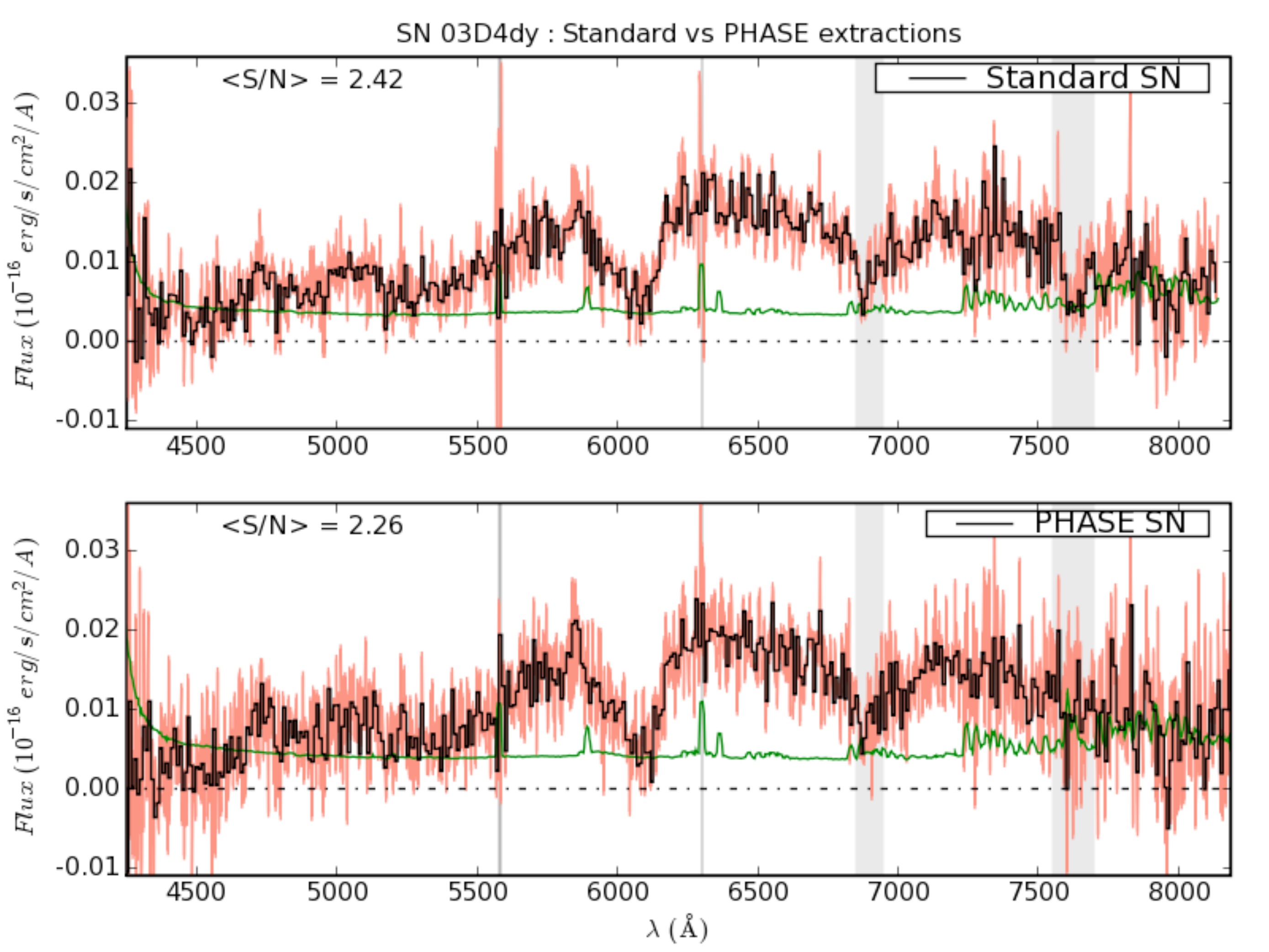}}\\[2mm]
\end{center}
\caption{Comparison of standard (top) and PHASE (bottom) extractions for the simple case of SN~03D4dy, an SN~Ia at $z=0.6$. Spectra have been rebinned with 10 $\AA$ (black) and are presented in the observer frame. The green spectra are the noise model for each
extraction. The statistical noise is plotted in red. The grey lines correspond to the intense atmospheric \ion{O}{[i]} emissions and the grey bands to atmospheric $O_2$ absorptions.}
\label{fig:CompFavorable03D4dy}
\vspace*{3cm}
\end{figure*}

We compare here the spectra obtained with the standard
and PHASE techniques, in the nearby subset 
of our sample, where the supernova is resolved from its host.
 This is the simplest possible situation and we expect that
the two extraction methods yield comparable results. Figure \ref{fig:CompFavorable03D4dy} shows the two extractions for SN~03D4dy,  an \Ia at $z=0.6$. The 
standard extraction is shown in the top panel while the PHASE extraction
is presented in the bottom panel. The green spectra are the noise model for each extraction.  The host
is not visible in the slit and the model is a simple PSF Gaussian of width equal to the seeing. As expected in this simple case, the two extractions yield 
comparable results in the
whole spectral range. To further compare the two extracted spectra, we define 
the quantity $<S/N>$ as the signal-to-noise ratio per pixel, involving the statistical noise $N$ propagated during the processing (PHASE or standard) and averaged over all pixels.
In the present example, we find that $<S/N>$ is slightly higher for the
standard extraction ($<S/N>=2.42$) than for the PHASE extraction ($<S/N>$=2.26).

\subsubsection{SN~03D4ag: a supernova in the arm of its spiral host}

We consider here the case of SN~03D4ag, an \Ia that exploded in the arm of a spiral galaxy at $z=0.285$. The definition of the sources is of critical importance in this case. In the standard procedure, two extractions are performed, one at the location of the supernova, the other in a region where the host signal is not contaminated by the SN. Then, this ``pure" host spectrum is subtracted from the other to get the supernova spectrum. This subtraction affects the colour of the final supernova spectrum.

\begin{figure*}
\begin{center}
\resizebox{14cm}{!}{\includegraphics{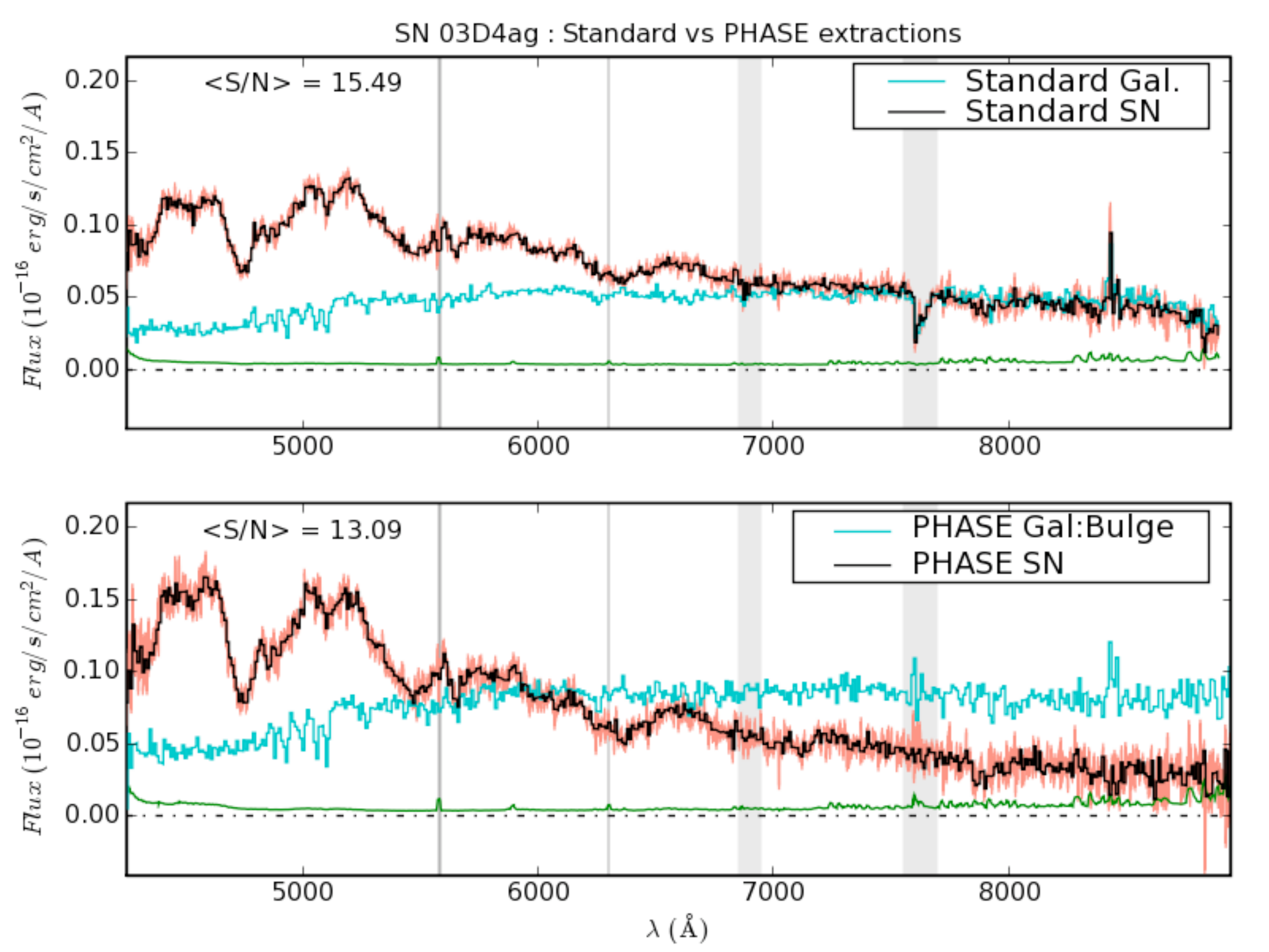}}\\[2mm]
\end{center}
\caption{Same as Fig. \ref{fig:CompFavorable03D4dy} for SN~03D4ag, an \Ia that exploded in the arm of a spiral galaxy at $z=0.285$. The Mix host model presented in the left panel of Fig. 
\ref{fig:PhaseModel} is used to recover the different components. The PHASE supernova spectrum is bluer than the standard spectrum, as it should be for this
$\phi \approx -8$ days \Ia.}
\vfill
\label{fig:CompDifficult03D4ag}
\end{figure*}

With PHASE, all sources are extracted at the same time provided that the sources are properly modelled. Figure \ref{fig:CompDifficult03D4ag} shows the comparison between the two extractions. The standard extraction limits the size of the
extraction window to prevent from too high host contamination and only part of the SN flux is extracted. On the contrary, if the host is correctly modelled, PHASE assigns the correct flux to the supernova (see Section \ref{sec:simul} above). 

We find comparable $<S/N>$ for the two extractions (13.1 for PHASE and
15.5 for the standard extraction). The supernova spectrum extracted with PHASE is blue, consistent with its early phase ($\phi \approx$ -8 days). This is
not the case for the standard spectrum that looks like an \Ia spectrum
around maximum light.

\subsubsection{SN~04D2bt: a supernova close to its host centre}

\begin{figure*}
\begin{center}
\resizebox{14cm}{!}{\includegraphics{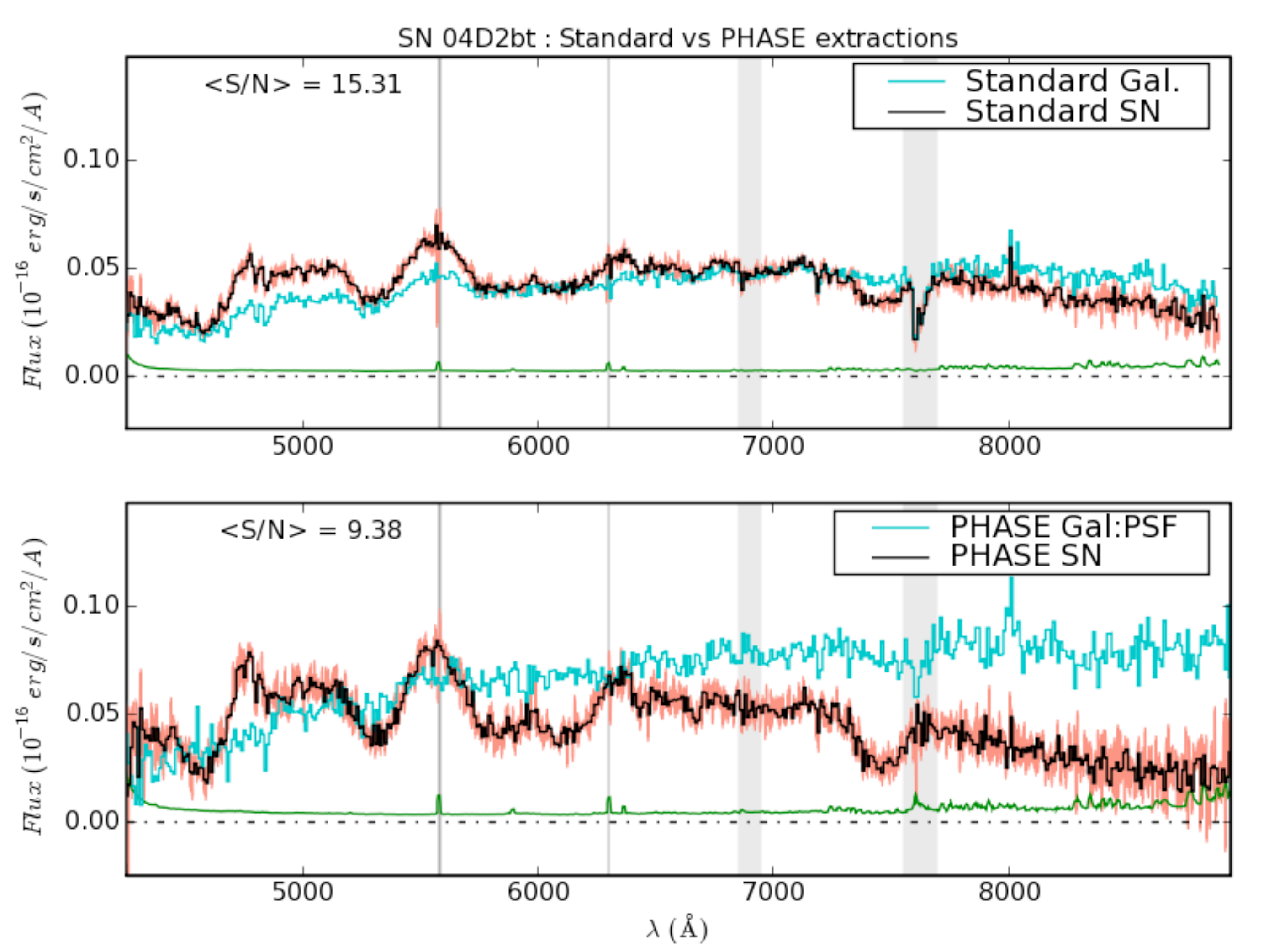}}\\[2mm]
\end{center}
\caption{Same as Fig. \ref{fig:CompFavorable03D4dy} for SN~04D2bt, an \Ia at $z=0.22$ that
exploded very close to its host centre, which renders the component separation difficult. Here, a PSF model is used for the host and permits a clean
restoration of the two components at the expense of a higher noise level.}
\vspace*{3cm}
\vfill
\label{fig:CompDifficult04D2bt}
\end{figure*}

SN~04D2bt is an \Ia at $z=0.22$ that exploded very close to its host centre 
(0.35$''$ for a seeing of 0.62$''$). This renders  the component separation 
difficult. Using a PSF model for the underlying galaxy, 
PHASE algorithm nevertheless 
succeeds in extracting separately 
the supernova and the host, as appears in Fig. \ref{fig:CompDifficult04D2bt}. 
The host and supernova spectra obtained from the standard procedure are strongly correlated and the supernova flux is thus not accurate. With PHASE, both components are recovered correctly at the expense of a higher noise level.

\section{Using SALT2 for the purpose of identifying SNLS spectra}

In this Section, we turn to the identification of SNLS supernovae. Once an optimal spectrum of a potential supernova (sometimes combined with its host signal, see above) has been extracted using the PHASE extraction method, it is essential to assess the type of the candidate and to determine its redshift for the Hubble diagram. 

\subsection{Redshift determination}

The redshift determination is based on the identification of a few galactic absorption or emission
lines (e.g., [\ion{O}{\sc ii}], [\ion{O}{\sc iii}], Hydrogen Balmer series, \ion{Ca}{H\&K}, $4000\, \AA$ break). A Gaussian fit of one or more of these lines
yields a typical uncertainty on the redshift of $\delta z\leq 0.001$ over the available spectral range. This is similar to the redshift determination accuracy obtained for similar spectral resolutions in \citet{Lidman05, Hook05, Howell05, Balland06, Balland07}. 

The separate extraction of the host signal from the SN in a vast majority of cases simplifies galaxy line identification and redshift determination. However, in about 20\% of cases, 
no host signal is visible at all and redshift determination has to be performed on SN features, from a series of fits of the supernova spectrum using the
SALT2 model (\citealt{Guy07}, see below). We allow the redshift to vary
in a plausible range, given the supernova features, with a step of $\delta z=0.005$ and we determine the redshift from the best-fit solution. This latter determination yields a typical uncertainty of $\delta z\approx 0.01$, ten times larger than for an
identification from host lines. This is in agreement with redshift determinations using correlation with a template, e.g. the SNID code developed by \citet{Blondin07}.  

\subsection{Identification}

The identification proceeds in two steps:

1) We systematically perform a simultaneous
fit of photometric (light curves) and spectroscopic data, using SALT2 
with and without a galaxy template (see below). This yields a set of
best-fit parameters whose values help at discriminating spectra with properties
different from the average properties of the SALT2 SNe~Ia training set \citep{Guy07}.\\

2) We then decide the type of the spectrum. 
This step is not automatic. We use both visual inspection of the
fitted spectrum and the parameter values of the best-fit SALT2 model. 
The final decision
is based on qualitative grounds. Nevertheless, using the SALT2 fit greatly 
helps in discriminating SNe~Ia from other types, as both spectroscopic and 
photometric fit parameters are different for the two. In particular,
we know the date of maximum, and thus the phase. Using the SALT2 fit 
allows us to alleviate the phase degeneracy between SNe~Ia and SNe~Ic (see below).\\

In the following, we detail these two steps.

The spectrophotometric template SALT2 \citep{Guy07} has been designed to fit the light curves of observed supernovae. As it is trained against a whole set of \Iae spectra templates, including both local, published \Iae templates and more distant \Iae spectra, including 39 SNLS first year spectra both from the VLT \citep{Balland08} and Gemini \citep{Howell05} telescopes, it is possible to use it for
a combined fit of the photometric data and the spectrum of a given supernova. This procedure has been briefly discussed in \citet{Guy07} and is extensively used in the present work. It offers the great advantage over a simple spectral fitting as the ones used in a lot of previous works concerned with \Iae spectral identification \citep{Lidman05, Matheson05, Foley08, Balland06, Balland07, Hook05, Howell05, Wood-Vasey07} to explicitly incorporate, in the spectral identification process, information on the date of B-band maximum light and colour. This, in particular, alleviates the possibility of confusing a Type Ib/c  supernova (SN~Ib/c) with an SN~Ia, at least at early phases, and has yielded, in several cases, a firm type determination of dubious spectra. It is also a good way to discriminate Branch-normal \Iae \citep{Branch93a,Branch06} from more peculiar objects for which SALT2 fails at providing a satisfying spectrophotometric model.

The identification using SALT2 is based on a $\chi^2$ minimisation procedure from the combined fitting of photometric and spectral data for a given supernova. The main photometric parameters entering the fit are the principal components $x_0$, $x_1$ of the SALT2 model (see \citealt{Guy07} for details) and the colour index $c$, defined as the difference between $(B-V)_{SN}$ and the average $<B-V>$ value for the whole training sample. The colour law of the model is obtained during the training process on the SALT2 training sample, along with the model components (see Fig. 3 of \citealt{Guy07} for a comparison to a \citealt{cardelli89} law).
The $x_1$ parameter is linked to the stretch of the supernova and can be interpreted as the number of standard deviations the stretch of the supernova under study is with respect to the whole training set. A set of spectroscopic parameters are fitted along with the photometric parameters and include a possible host fraction, an overall normalisation parameter and a tilt parameter. These two latter parameters allow us to take into account possible errors in the flux calibration, due to, e.g., variations from the time average response function used for flux calibration, or differential refraction effects improperly corrected for. Higher order re-calibration parameters can be added at will, but we usually limit the possibility of spectral re-calibration to the two parameters described above, except for high S/N spectra, for which adding a third re-calibration parameter improves the overall fit by
also correcting for ``curvature" effect along the spectrum. 

\begin{figure*}[htbp]
\begin{center}

\begin{tabular}[b]{cc}
  \hspace{-5mm}
  \vspace*{1cm}
  \begin{minipage}[b]{7.5cm}
    \centering
    \fbox{\includegraphics[width=6cm,height=2cm]{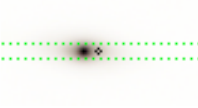}}
  \end{minipage}
  &
  \begin{minipage}[b]{7.5cm}
  \end{minipage}
  \\
  \hspace{-5mm}
  \includegraphics[width=8.5cm]{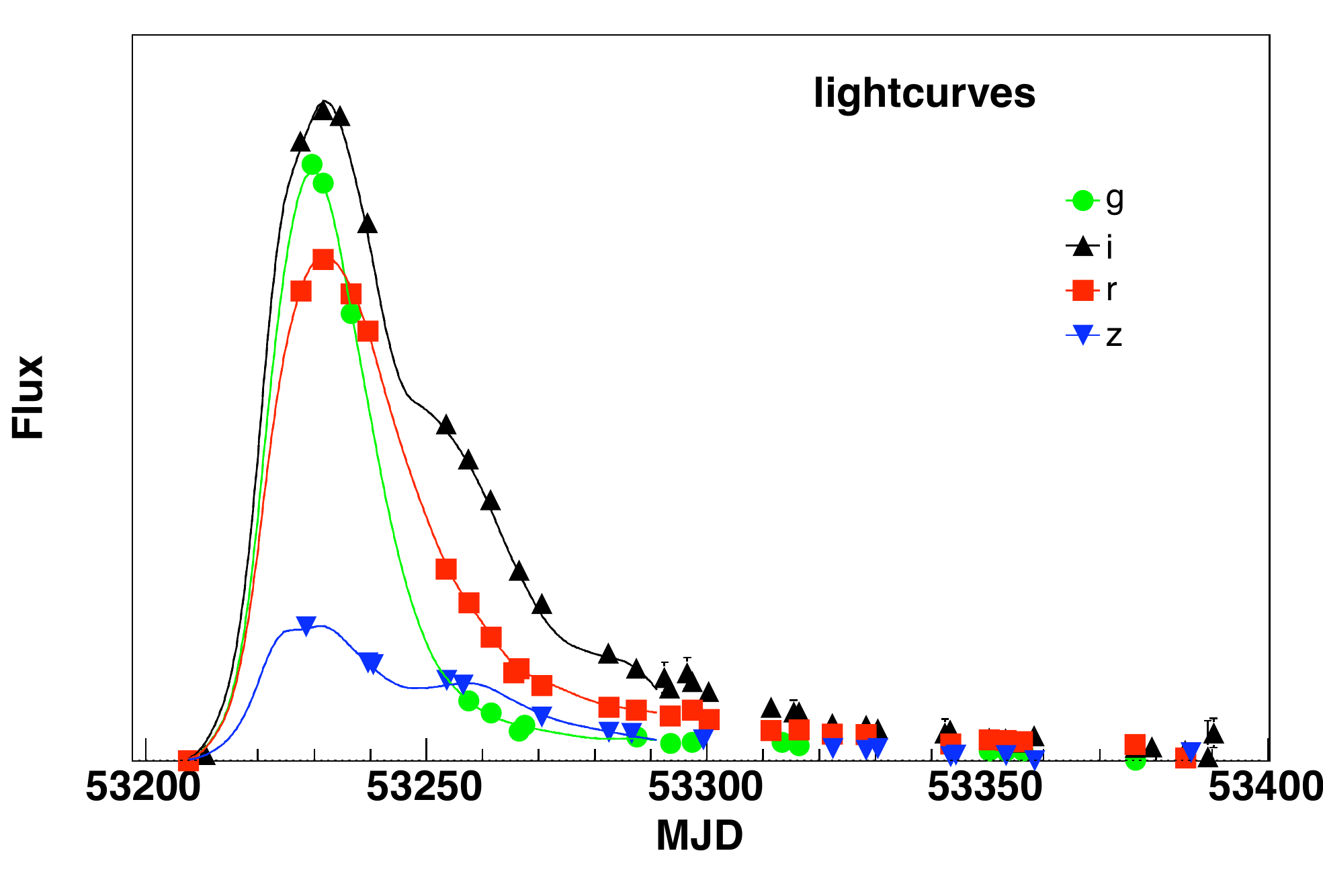}
  &
  \vspace*{-3mm}\includegraphics[width=8.8cm]{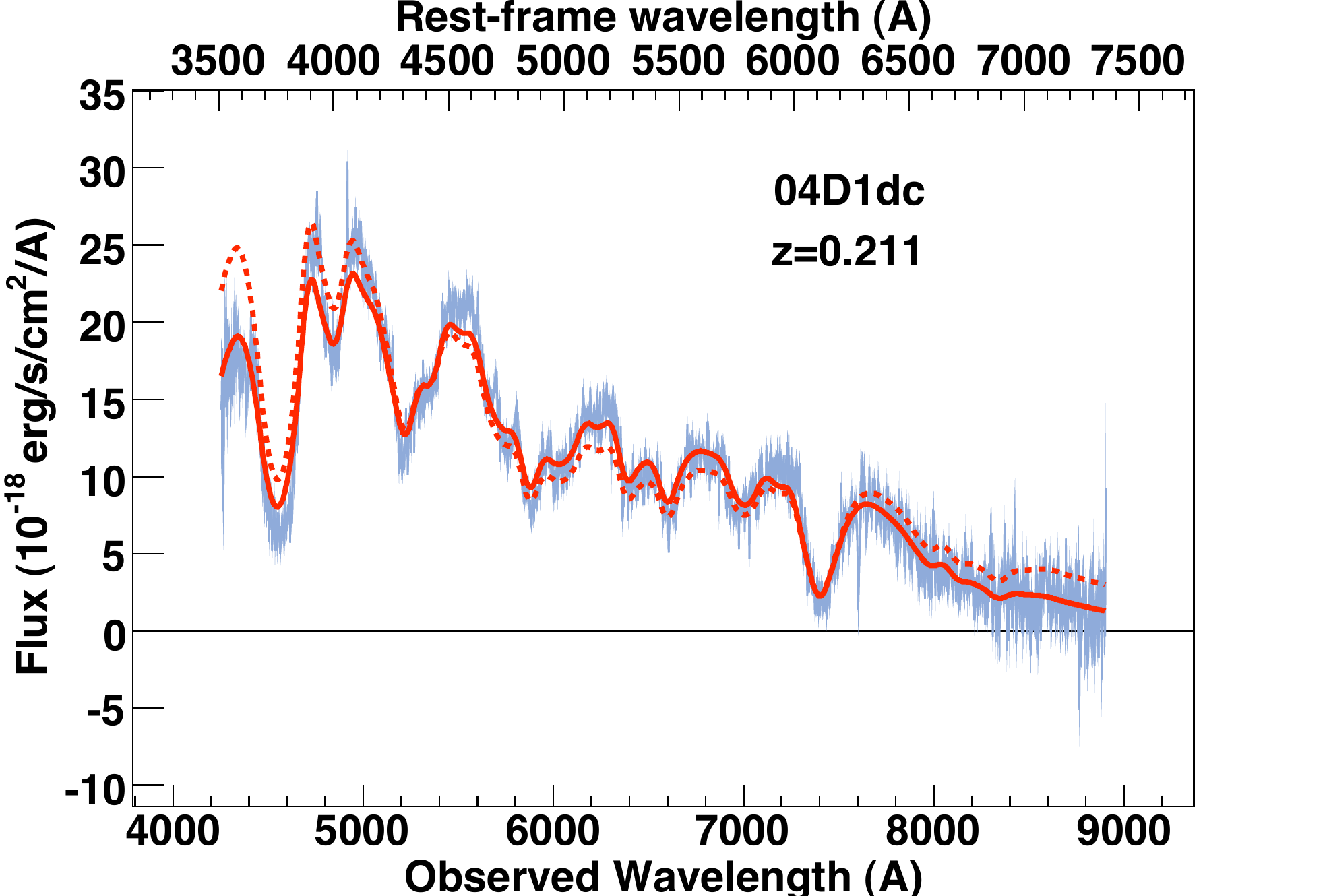}\\
\end{tabular}
\caption{SALT2 fit of SN~04D1dc, an \Ia spectroscopically observed at maximum light. Top: negative deep reference image of SN~04D1dc host. The locus of the SN explosion is indicated with a ``+" sign. The position of the slit is shown as green dotted lines. Bottom left: restframe $g_M$,$r_M$,$i_M$,$z_M$ light curves and corresponding SALT2 fits. Bottom right: 
PHASE extracted spectrum with the best-fit SALT2 model overlapped. Dashed red line is the raw SALT2 model. Solid line shows the best-fit model after re-calibration using 3 parameters. The spectrum is presented in the observer frame and the restframe wavelength is reported on top.
  \label{fig:04D1dc}
}
\end{center}
\end{figure*}

Figure \ref{fig:04D1dc} presents an example of the combined identification performed with SALT2. Recall that SN~04D1dc is an \Ia at maximum light at $z=0.211$, well separated from its host core (see top panel of Fig. \ref{fig:04D1dc}). 
A separate PHASE extraction of the host and the supernova was possible, yielding an almost host-free supernova spectrum (Section \ref{sec:speccalib} and Fig. \ref{fig:PSFspec}). 
The bottom left panel of Fig. \ref{fig:04D1dc} shows the $g_M$, $r_M$, $i_M$, and $z_M$ photometric data, along with the  
best-fit SALT2 light curves models overlapped. The fit is excellent in each band and the secondary maxima, typical of SNe~Ia, are clearly seen in the infrared light curves. The bottom right panel of Fig. \ref{fig:04D1dc} shows the 
corresponding spectral fit. The dashed red line is the SALT2 model obtained 
with no re-calibration, the solid red line being the same model but 
with re-calibration (using three re-calibration parameters in this specific case). In this example, re-calibration 
appears to have a weak effect on the final fit (significant only around 4000-4500 $\AA$).
This indicates that flux calibration is accurate. Strong discrepancies between the dashed and solid red lines appear in some cases, hinting either towards a problem in flux calibration, or, more probably, an inadequacy of the model to reproduce the spectral data. This happens in particular for spectra of supernovae that cannot be adequately reproduced by SALT2 (such as SNe~Ic) as
only SNe~Ia are considered in the SALT2 training sample.

When it was not possible to extract separately the supernova from the host signal with PHASE, the SALT2 fit incorporates a galaxy template in the model \citep{Guy07}. We use galaxy spectral series synthesised by PEGASE2 \citep{Fioc97,Fioc99} for various Hubble types (Elliptical, S0, Sa, Sb, Sbc, Sc and Sd). Each series ranges from 1 Gyr to 13 Gyrs and offers a continuous sequence from blue to red within each Hubble type. We also use the spectral templates for Elliptical, S0, Sa, Sb and Sc types from \citet{Kinney96}. The best-fit interpolated galaxy template and age are obtained as part of the minimising procedure.

We have checked the quality of PHASE SN/Host separation. To do this, we have selected all
the \Iae of the 1st VLT large programme for which a
separate extraction of the SN component from its host was
possible. We then did a SALT2 fit of these supernovae
with a galactic component in the fitted model.
For this specific test, we only used the \citet{Kinney96} Hubble
sequence (from Elliptical to Sc types). We discarded four extinguished 
supernovae with very red colours and two supernovae for which no acceptable
fit was obtained. We ended up with a sample of 68 \Iae (among which 47 were identified as certain SNe~Ia, see our classification definition below).
With the whole sample, we find that the average galaxy fraction in the best-fit model is
$4.5\pm 3.0$ \% with a dispersion of 25\%. If we restrict ourselves
to the sample of 47 certain SNe~Ia, we find an average galaxy 
fraction of $0.7\pm 3.2$ \% and the dispersion slightly decreases to 22 \%.
These results confirm that residual host light contamination of the PHASE extracted supernovae is moderate.

For each SNLS supernova for which both photometric data (in at least two bands) and a spectrum are available, we perform SALT2 fits using all galaxy types with two re-calibration parameters. Light curves are fitted in the range -15 and +40 days, which turns out to be sufficient for most supernovae under investigation. Spectra are fitted over their whole spectral range. The best-fitting model is for the minimum combined (photometric and spectral) $\chi^2$ value. Result parameters and fits are inspected to assess the identification of the candidate as a \Ia (step 2 of our identification process). Large $x_1$ absolute values hint possible peculiarities (under or over-luminous SN). Large positive colour indexes are either due to large intrinsic or galactic  reddenings, or possible SN~Ib/c. Large negative values are found for Type II supernovae (SN~II), for which  no good spectral fits are obtained. Light curves and spectral fits are visually inspected. When strong re-calibration is necessary to obtain a good spectral fit, we track down the reason for this. Often, this is the sign of a non \Ia spectrum.  The combination of all these indicators usually leads to an optimised identification, even though the final decision relies on human judgement.

Spectra of \Iae are classified according to the following scheme: SN~Ia (certain SN~Ia), SN~Ia$\star$ (possible SN~Ia but other types, in particular SN~Ic, cannot be excluded given the S/N and phase), SN~Ia\_pec (peculiar SN~Ia), SN (certain SN as the spectrophotometric properties match the known properties of an SN although the type is not clear), SN? (possible SN of unclear type, other objects such as variable stars or AGNs are possible), SN~Ib/c, SN~II. Note that this classification scheme slightly 
differs from the classification adopted in \citet{Astier06}. The Hubble diagram is built only from \Ia and SN~Ia$\star$ supernovae or from \Ia supernovae alone, underlying the necessity for a clean supernova identification.

\subsection{Results of the identification}

In this Section, we illustrate the performances of our SALT2-based identification through a series of representative examples. The bulk of our \Iae
identifications will be published in \citet{Balland08}. In the examples given below, we show the results of the spectral and light curve fits for the
sake of completeness. Spectra are presented in the observer frame and the restframe wavelength is reported on top of the corresponding Figure. The SALT2 model output parameters are summarised for each example in Table \ref{tab1}.

\subsubsection{SN~05D2dt: a high S/N SN~Ia at $z\sim 0.6$}

\begin{figure*}[htbp]
\begin{center}

\begin{tabular}[b]{cc}
  \hspace{-5mm}
  \vspace*{1cm}
  \begin{minipage}[b]{7.5cm}
    \centering
    \fbox{\includegraphics[width=6cm,height=2cm]{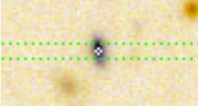}}
  \end{minipage}
  &
  \begin{minipage}[b]{7.5cm}
    \hspace{-5mm}\includegraphics[width=8cm]{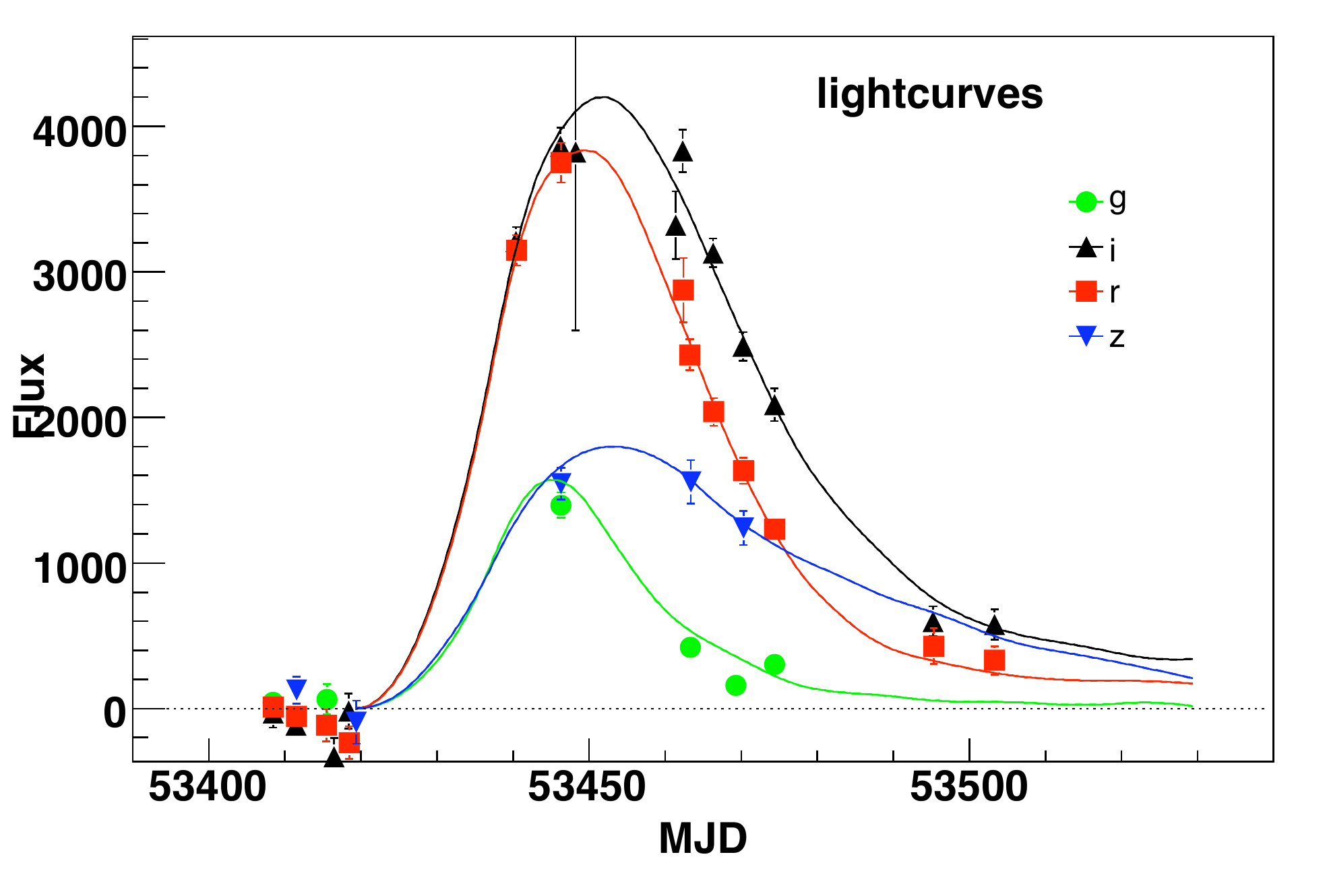}
  \end{minipage}
  \\
  \hspace{-5mm}
  \includegraphics[width=8.5cm]{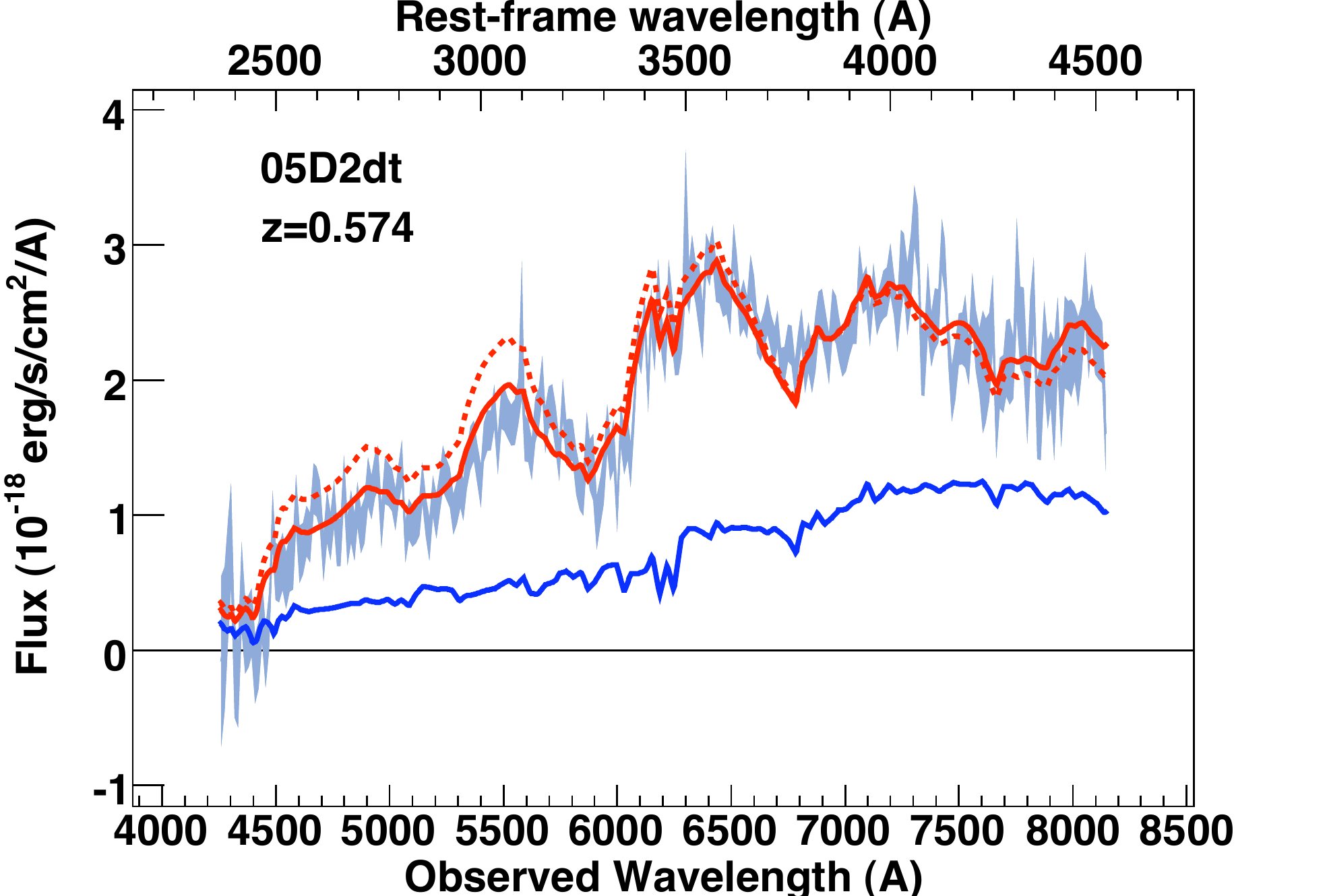}
  &
  \vspace*{-3mm}\includegraphics[width=8.8cm]{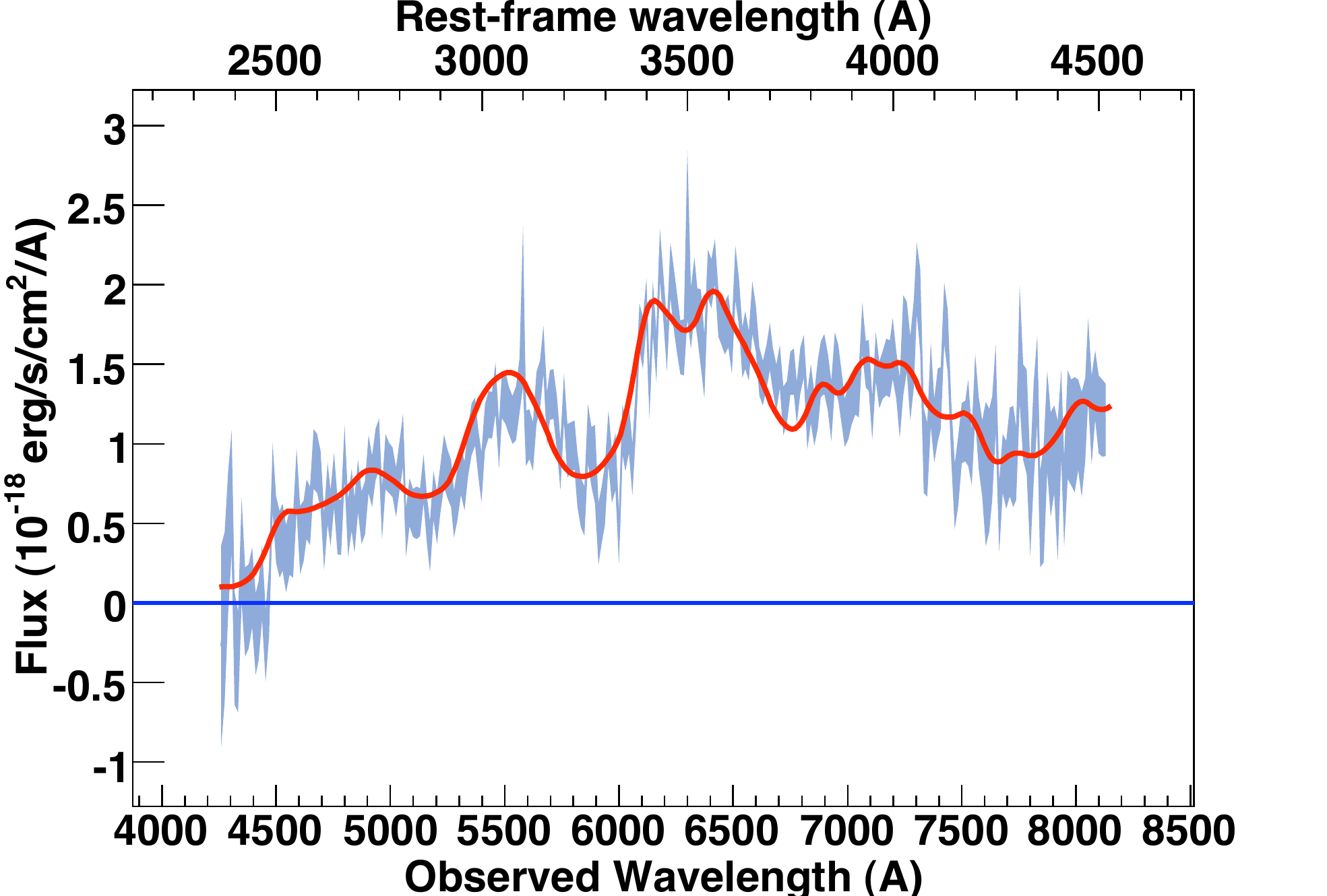}\\
\end{tabular}
\caption{SALT2 fit of SN~05D2dt. Top left: negative deep reference image of SN~05D2dt. The locus of the SN is indicated as a ``+" sign. The slit position is shown as green dotted lines. Top right: restframe $g_M$,$r_M$,$i_M$,$z_M$ light curves and corresponding SALT2 fits. Bottom left: full PHASE extracted spectrum (including host signal) with SALT2 raw (dashed line) and re-calibrated (solid line) models. 
For the latter, two re-calibration parameters have been used. Bottom right: host-subtracted spectrum with SALT2 best-fit model overlapped. The spectra have been rebinned to 15 \AA. This supernova is identified as an SN~Ia.
  \label{fig:05D2dt}
}
\end{center}
\end{figure*}

In Fig. \ref{fig:04D1dc}, we have shown the case of a host free, low-redshift
\Ia at maximum light. 
Figure \ref{fig:05D2dt} shows the spectrum of SN~05D2dt, a $\phi \approx  -2$ days \Ia 
at $z=0.574$ (about the average redshift of the SNLS sample),  
observed at the VLT. This supernova lies at the centre of its host (see top left panel of Fig. \ref{fig:05D2dt}) and no 
separate extraction was possible. The bottom left panel of Fig. \ref{fig:05D2dt} shows the full, 
host-contaminated spectrum and the SALT2 best-fit model (before and after 
re-calibration). The bottom right panel shows the host-subtracted spectrum, with the 
calibrated model overlapped. In both cases, the spectrum has been rebinned with 15 $\AA$ bins for the sake of visual convenience. In the bottom left panel, the model 
galaxy is plotted as a blue solid line. Contribution of the host model to the 
full model is computed in each photometric band. 
A ``bolometric" fraction of host is also evaluated over the full available 
spectral range and is given in Tab. \ref{tab1}. In the case of SN~05D2dt, a fair host contamination ($43\%$ of host template in the fitting model) is present 
and is well modelled by an early-type template. The model is tilted 
in order to adequately reproduce the spectrum, but this re-calibration is 
only moderate ($\approx 10$ \% flux re-calibration from $\lambda=4500$ \AA~ to $\lambda=8500$ $\AA$). Once the host contribution has been subtracted, the \Iae 
features are very well reproduced by the SALT2 model (bottom right panel of Fig. \ref{fig:05D2dt}) and the 
identification as an SN~Ia is unambiguous at this phase, even if no clear
\ion{Si}{\sc ii} can be seen around restframe 4000 $\AA\ $ due to the S/N level.

\subsubsection{SN~04D4ib: an \Ia buried into its host signal}

\begin{figure*}[htbp]
\begin{center}

\begin{tabular}[b]{cc}
  \hspace{-5mm}
  \vspace*{1cm}
  \begin{minipage}[b]{7.5cm}
    \centering
    \fbox{\includegraphics[width=6cm,height=2cm]{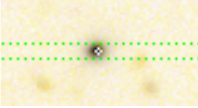}}
  \end{minipage}
  &
  \begin{minipage}[b]{7.5cm}
    \hspace{-5mm}\includegraphics[width=8cm]{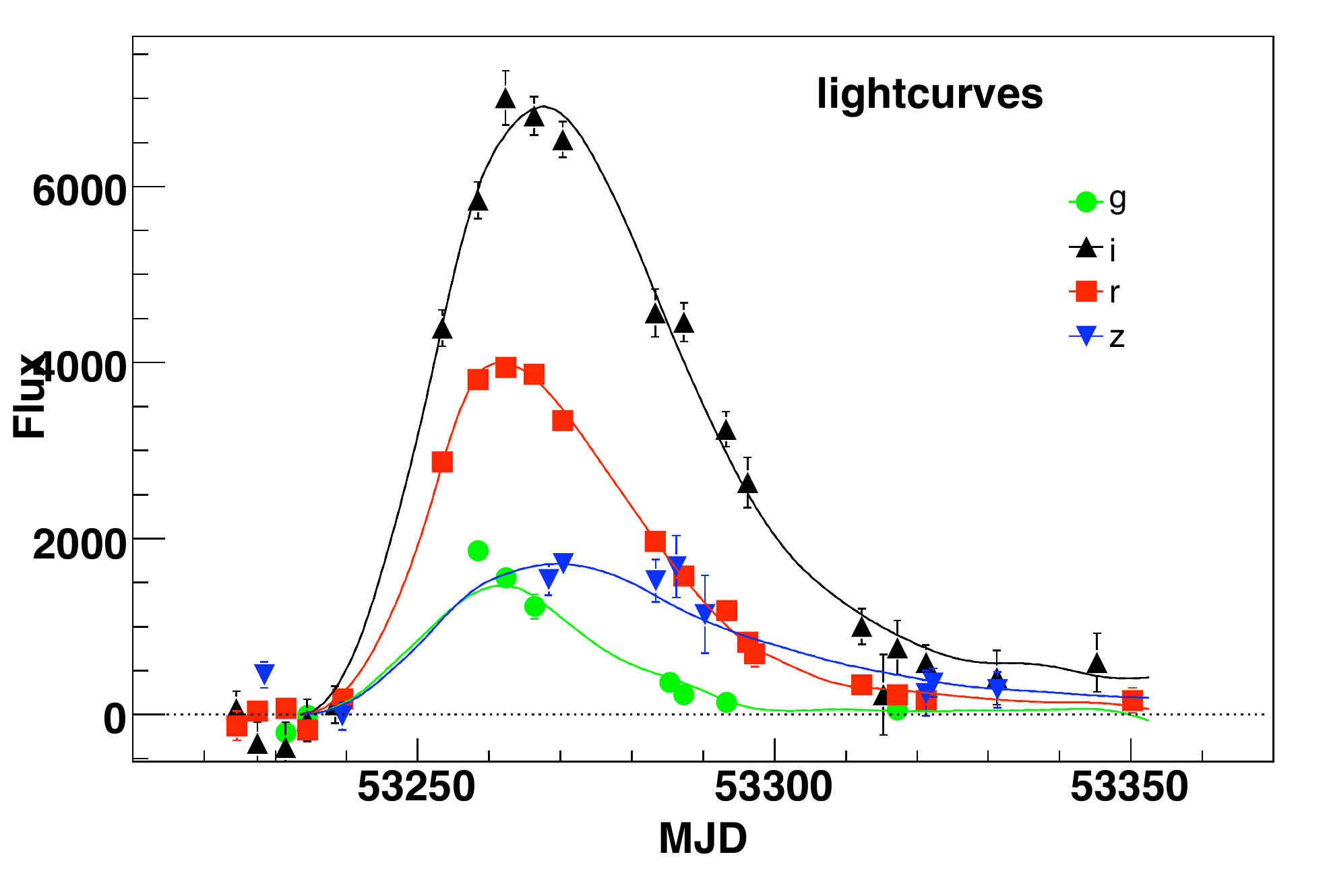}
  \end{minipage}
  \\
  \hspace{-5mm}
  \includegraphics[width=8.5cm]{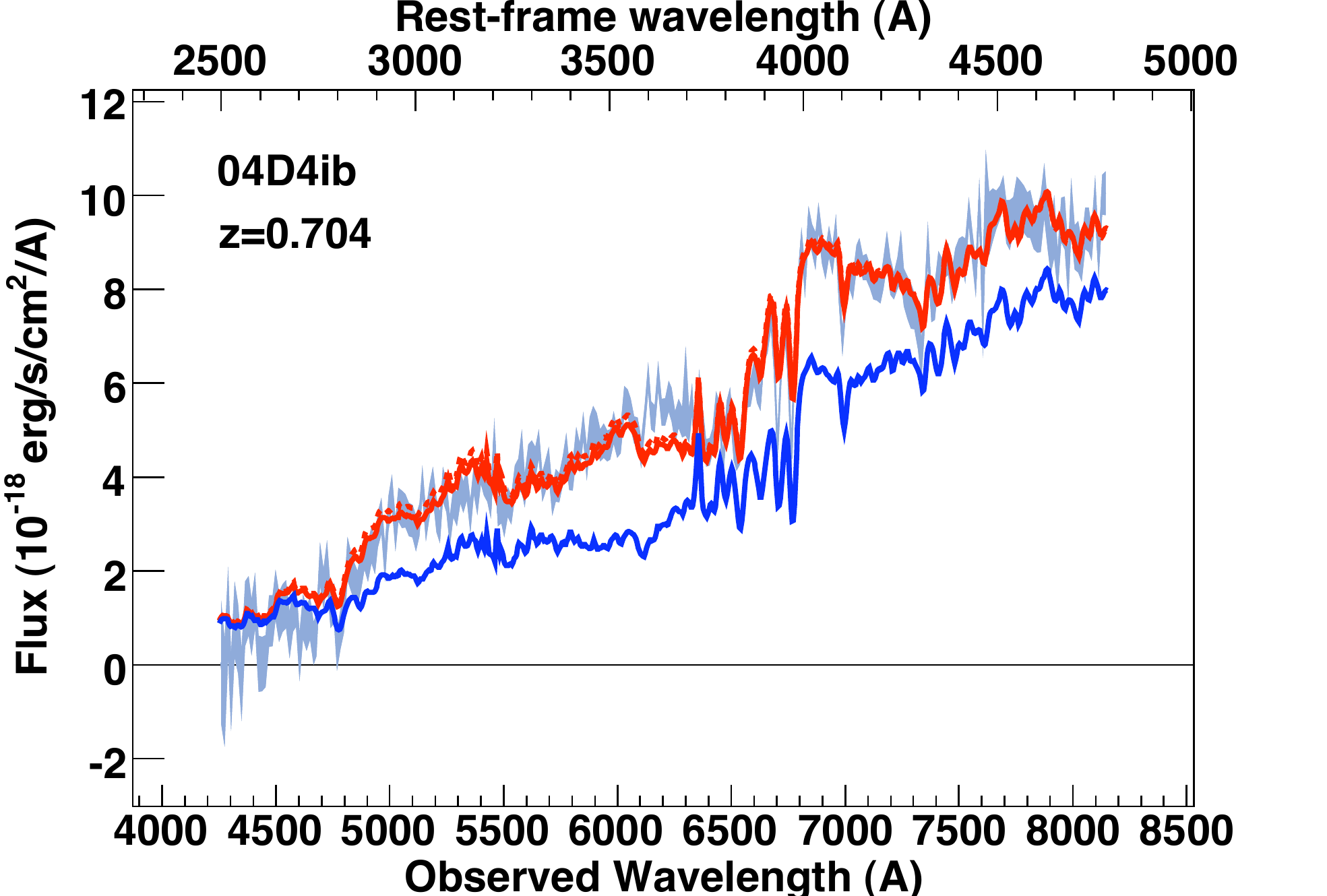}
  &
  \vspace*{-3mm}\includegraphics[width=8.8cm]{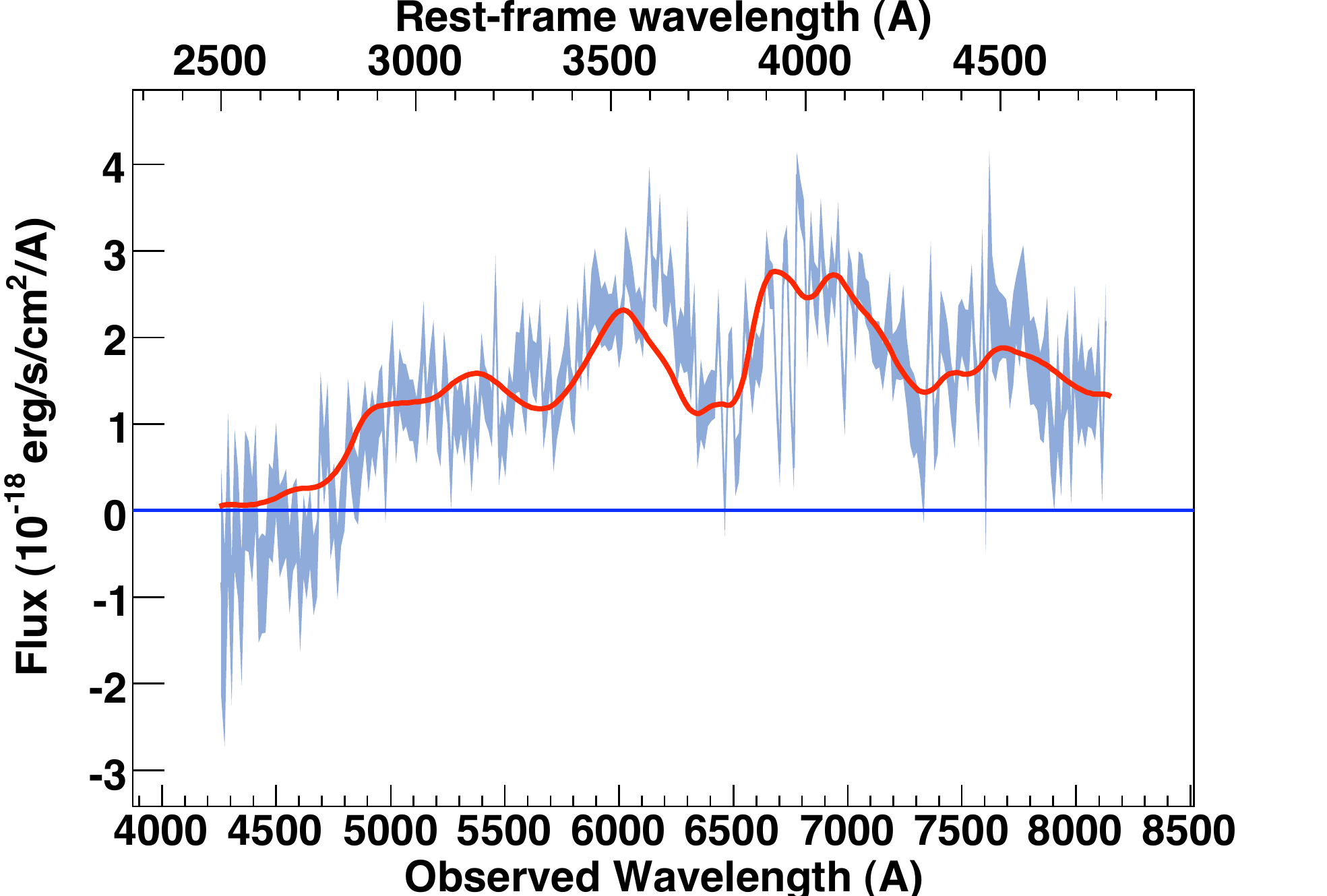}\\
\end{tabular}
\caption{Same as Fig. \ref{fig:05D2dt} for SN~04D4ib. This supernova is identified as an SN~Ia.
  \label{fig:04D4ib}
}
\end{center}
\end{figure*}

We show here SN~04D4ib, an \Ia at $z=0.704$ deeply 
contaminated by its host core signal. This is a ``{\em SNGAL}" case, for which 
PHASE cannot extract separately the supernova and the host. Figure \ref{fig:04D4ib} shows that the SALT2 fit is efficient at recovering the supernova spectrum (bottom right panel) from the full, host contaminated spectrum (bottom left panel). Here, a fraction of 71\% of a \citet{Kinney96} S0 template has been subtracted. Host \ion{Ca}{\sc ii} lines subtraction residuals are clearly seen in the supernova spectrum, but the broad SN features are well reproduced by the fit. No \ion{Si}{\sc ii}
 at 4000 \AA~ is visible, but given the phase of this SN ($\phi=+1$ day), confusion with an SN~Ic is unlikely (see below Section \ref{sec:CompIaIc}) and we can secure the type as an SN~Ia.

\subsubsection{SN~04D4dw: a distant ($z\sim 1$) SN~Ia$\star$}

\begin{figure*}[htbp]
\begin{center}

\begin{tabular}[b]{cc}
  \hspace{-5mm}
  \vspace*{1cm}
  \begin{minipage}[b]{7.5cm}
    \centering
    \fbox{\includegraphics[width=6cm,height=2cm]{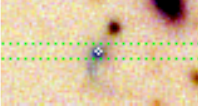}}
  \end{minipage}
  &
  \begin{minipage}[b]{7.5cm}
    \hspace{-5mm}\includegraphics[width=8cm]{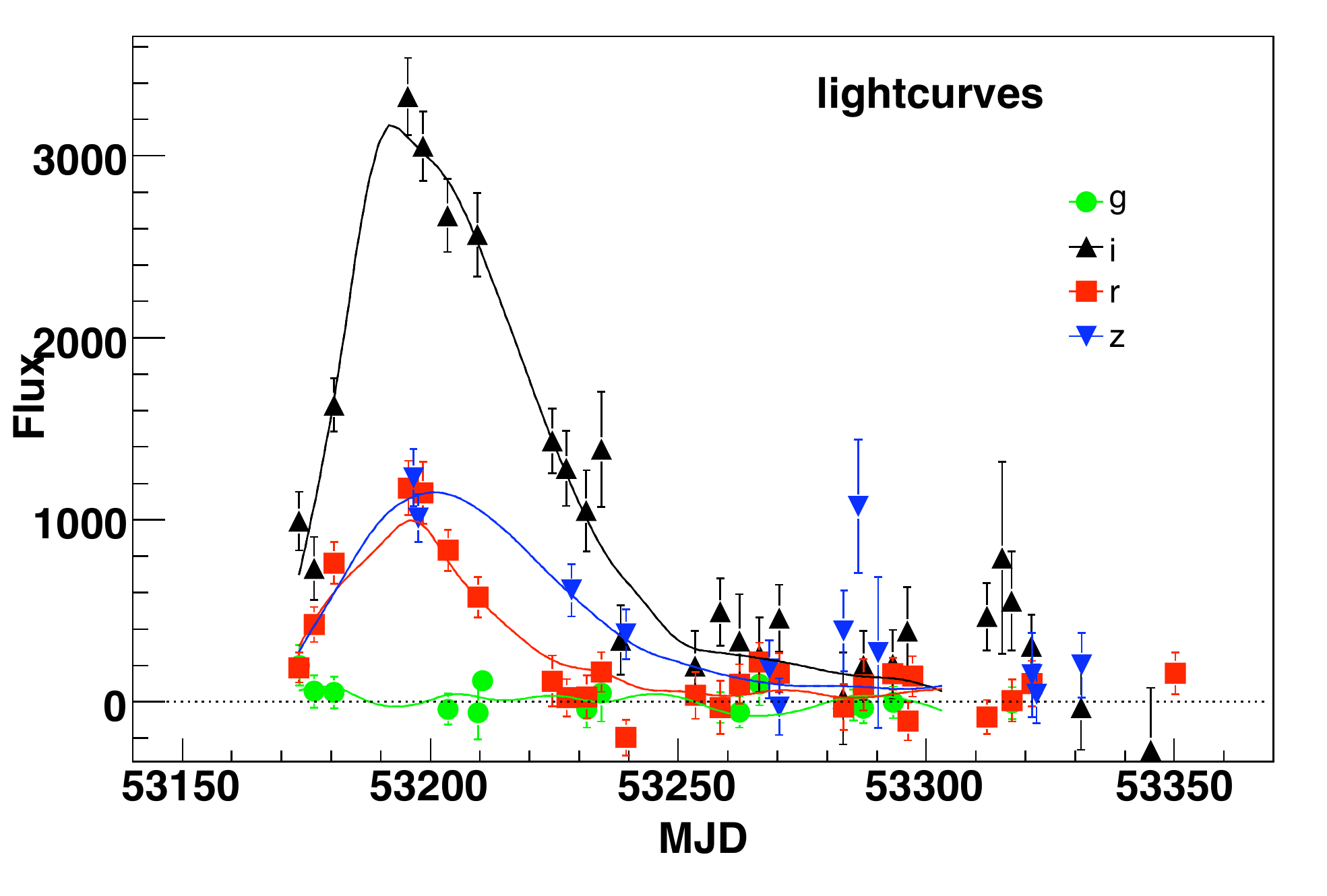}
  \end{minipage}
  \\
  \hspace{-5mm}
  \includegraphics[width=8.5cm]{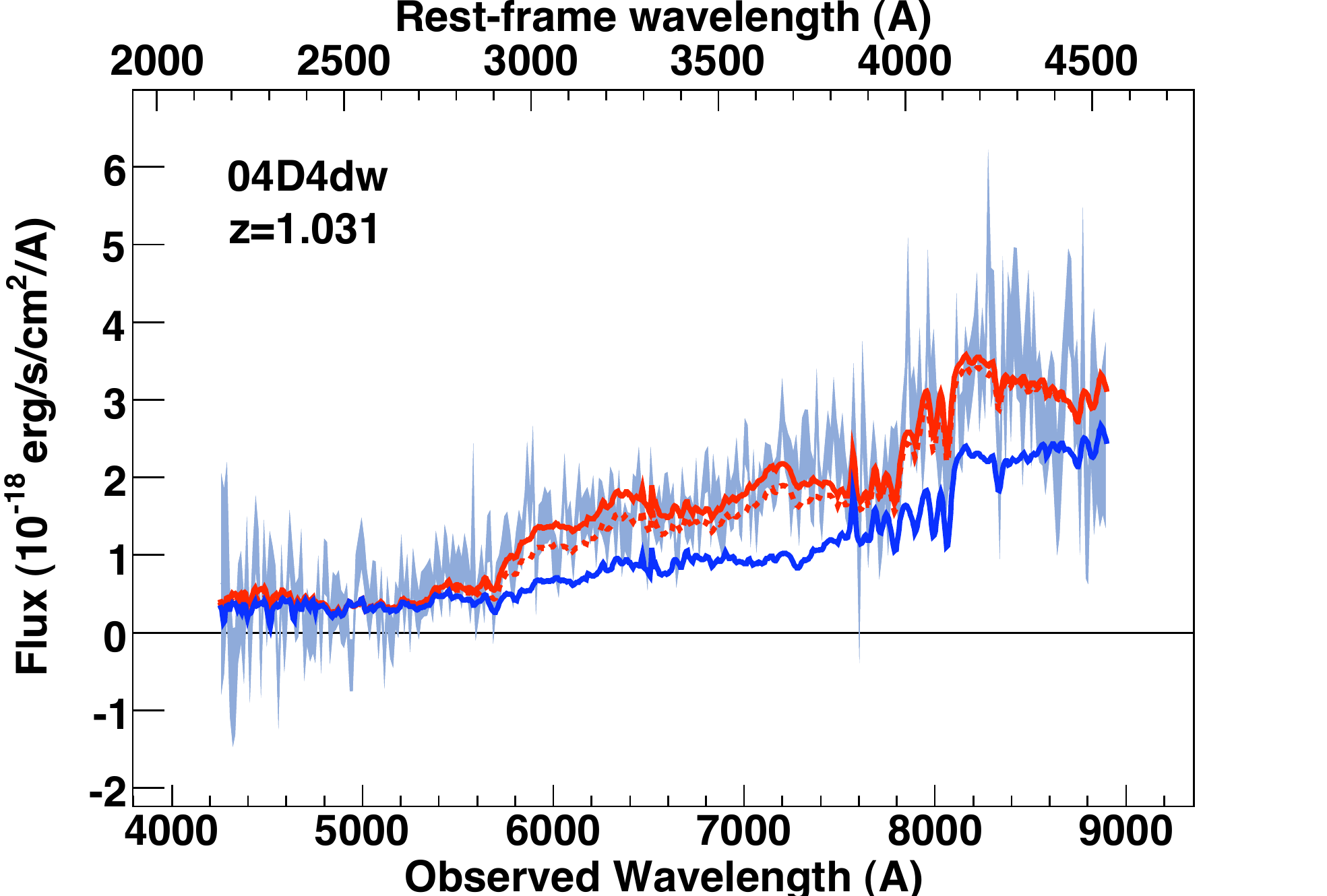}
  &
  \vspace*{-3mm}\includegraphics[width=8.8cm]{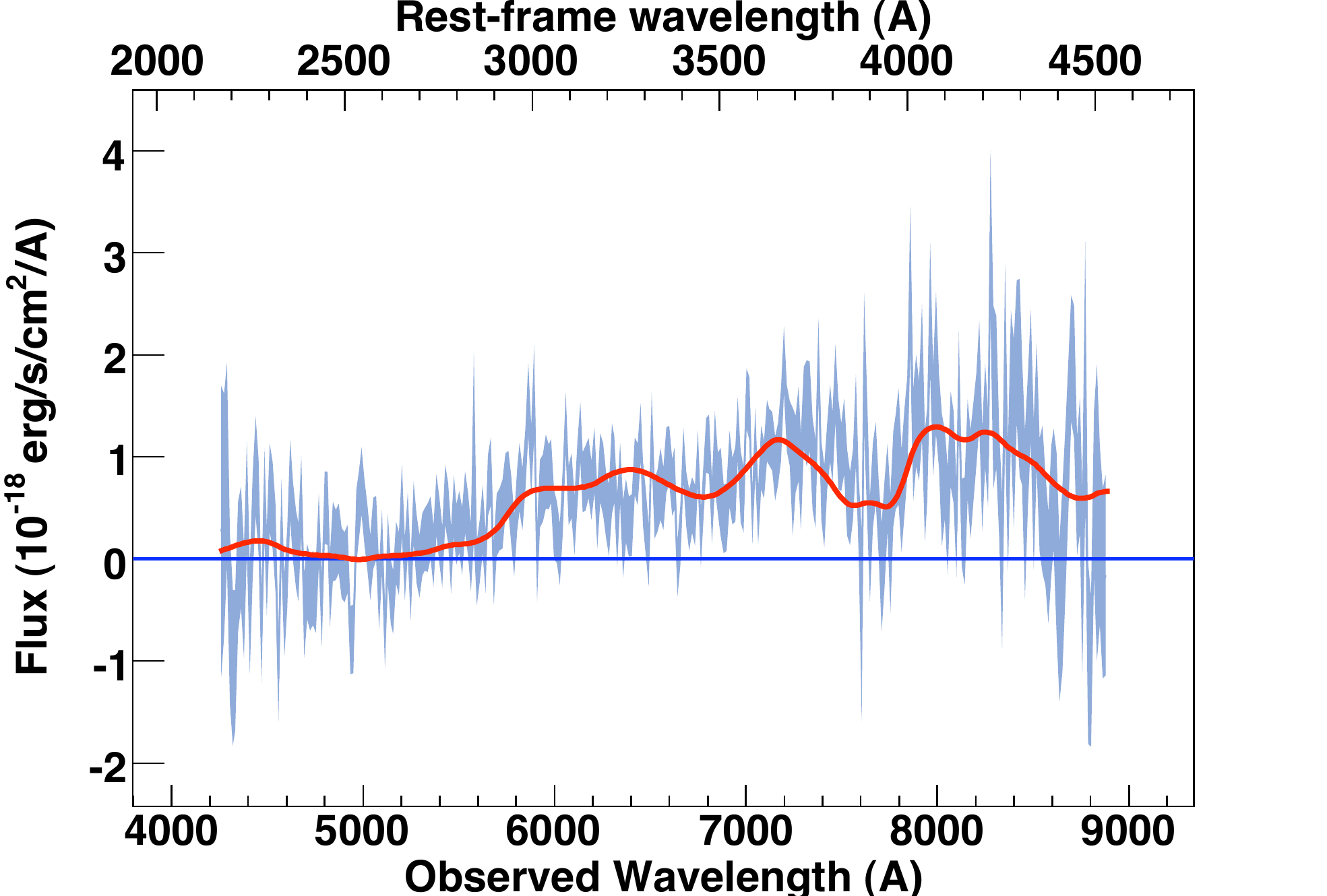}\\
\end{tabular}
\caption{Same as Fig. \ref{fig:05D2dt} for SN~04D4dw. This supernova is identified as an SN~Ia$\star$.
  \label{fig:04D4dw}
}
\end{center}
\end{figure*}

SN~04D4dw at $z=1.031$ is the second farthest \Ia of the 3rd year SNLS sample. This is another ``{\em SNGAL}" case for which PHASE is not able to efficiently perform a separate extraction. Figure \ref{fig:04D4dw} shows the full spectrum (bottom left panel) and host-subtracted supernova spectrum (bottom right panel). Due to its high redshift, the UV part of the spectrum is visible down to restframe 2100 \AA~ and is correctly reproduced by the SALT2 fit. This is not an obvious result, as the SALT2 training sample is rather poor in UV spectra \citep{Guy07}. 
The redshift is obtained from the presence of \ion{Ca}{\sc ii} absorption lines. 
The blueshifted \ion{Ca}{\sc ii} around restframe 3700 \AA~ and possibly some \ion{Si}{\sc ii} at restframe 4000 \AA~ are visible in the spectrum of this
supernova slightly past maximum ($\phi=2.1$ days). Given the rather low S/N, 
it is identified as an SN~Ia$\star$.

\subsubsection{SN~04D4jv: an SN~Ic supernova}
\label{sec:CompIaIc}
The possible confusion of an \Ia with an SN~Ic is a concern when identifying a supernova candidate. Indeed, the spectral structure of an \Ia slightly past maximum light resembles the one of an SN~Ic a few days before maximum (compare for instance the spectra of SN~2002bo, \citealt{Benetti04}, 4 days past maximum and
of SN~1994I, \citealt{Filippenko95}, 5 days before maximum). This degeneracy is amplified at larger redshift, as the spectral 
range of the observed spectrum and the S/N decrease with redshift and it is often difficult to confidently rule out the possibility of an SN~Ic. Using SALT2 for the purpose of identification 
alleviates this degeneracy, as the phase is in principle known from the light curve fitting. 
The fact that SALT2 fails at reproducing a spectrum is in itself an indication that it is unlikely to be an SN~Ia, as far as the phase and spectral coverage are well described by the SALT2 training set. Moreover, the values of the SALT2 colour value $c$ might also help, as SNe~Ic are redder than SNe~Ia at 
the same phase. Indeed, SNe~Ic tend to have high colour values, $c\geq 0.3$ (even if this sole result is not sufficient to unambiguously classify
a supernova as an SNe~Ic, as very extinguished \Iae would have comparable $c$ values). Nevertheless, it has been noted that SNe~Ic and \Iae spectra show
the greatest similarity about one week after maximum light \citep{Hook05,Howell05}.
In particular, iron absorption 
features are present in SNe~Ic spectra in the range $4200-5000$ \AA, that
resemble the spectral 
shape of \Iae $\sim$ 1-2 weeks after maximum light. 
Identification of later-phase spectra thus remains subject to caution, 
especially given
the S/N of SNLS spectra at $z\sim 0.5$. A substantial fraction of
supernovae classified as SN~Ia$\star$ in our sample correspond to this situation
of a $\sim 1$ week past maximum SN, for which a fair \Ia match is obtained 
with SALT2 but an SN~Ic solution cannot be
confidently ruled out. The remaining SN~Ia$\star$ are usually due to low
S/N. 

\begin{figure*}[htbp]
\begin{center}

\begin{tabular}[b]{cc}
  \hspace{-5mm}
  \vspace*{1cm}
  \begin{minipage}[b]{7.5cm}
    \centering
    \fbox{\includegraphics[width=6cm]{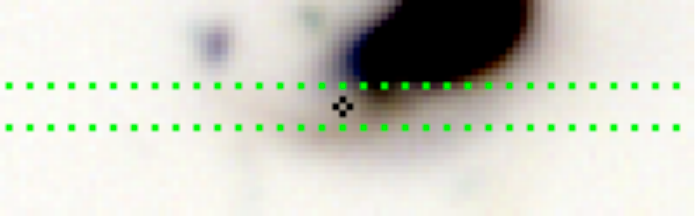}}
  \end{minipage}
  &
  \begin{minipage}[b]{7.5cm}
    \hspace{-5mm}\includegraphics[width=8cm]{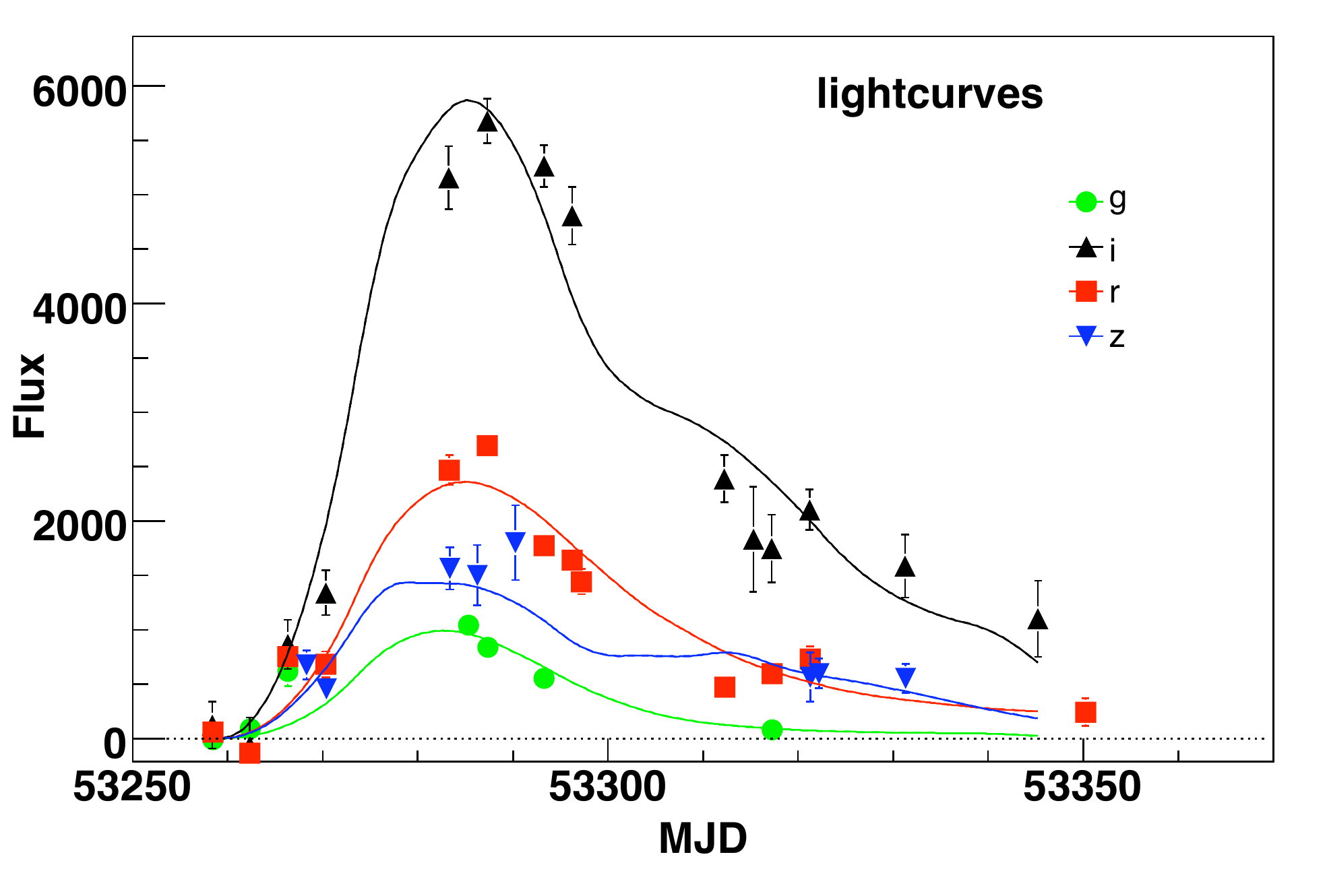}
  \end{minipage}
  \\
  \hspace{-5mm}
  \includegraphics[width=8.5cm]{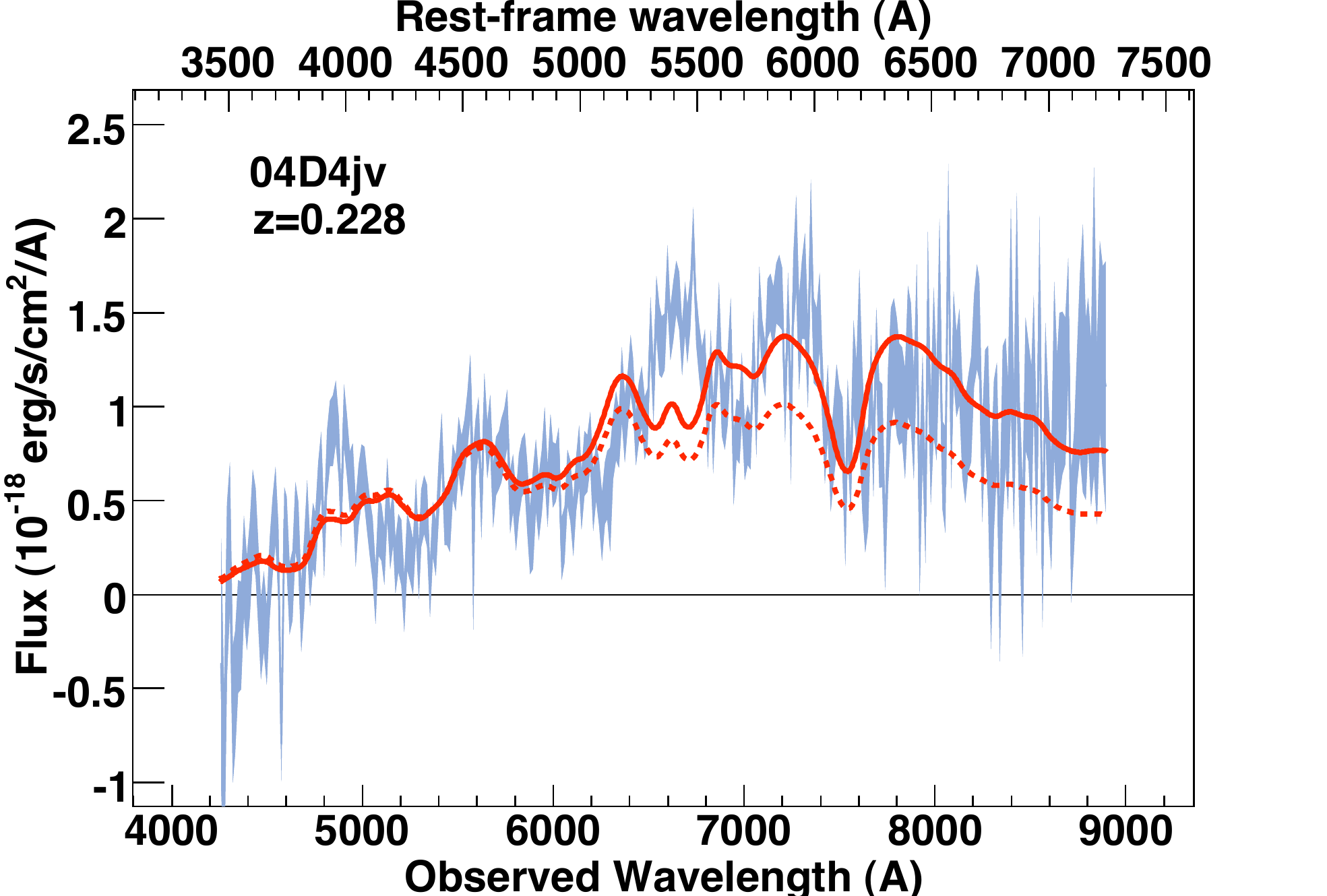}
  &
  \vspace*{-3mm}\includegraphics[width=8.8cm]{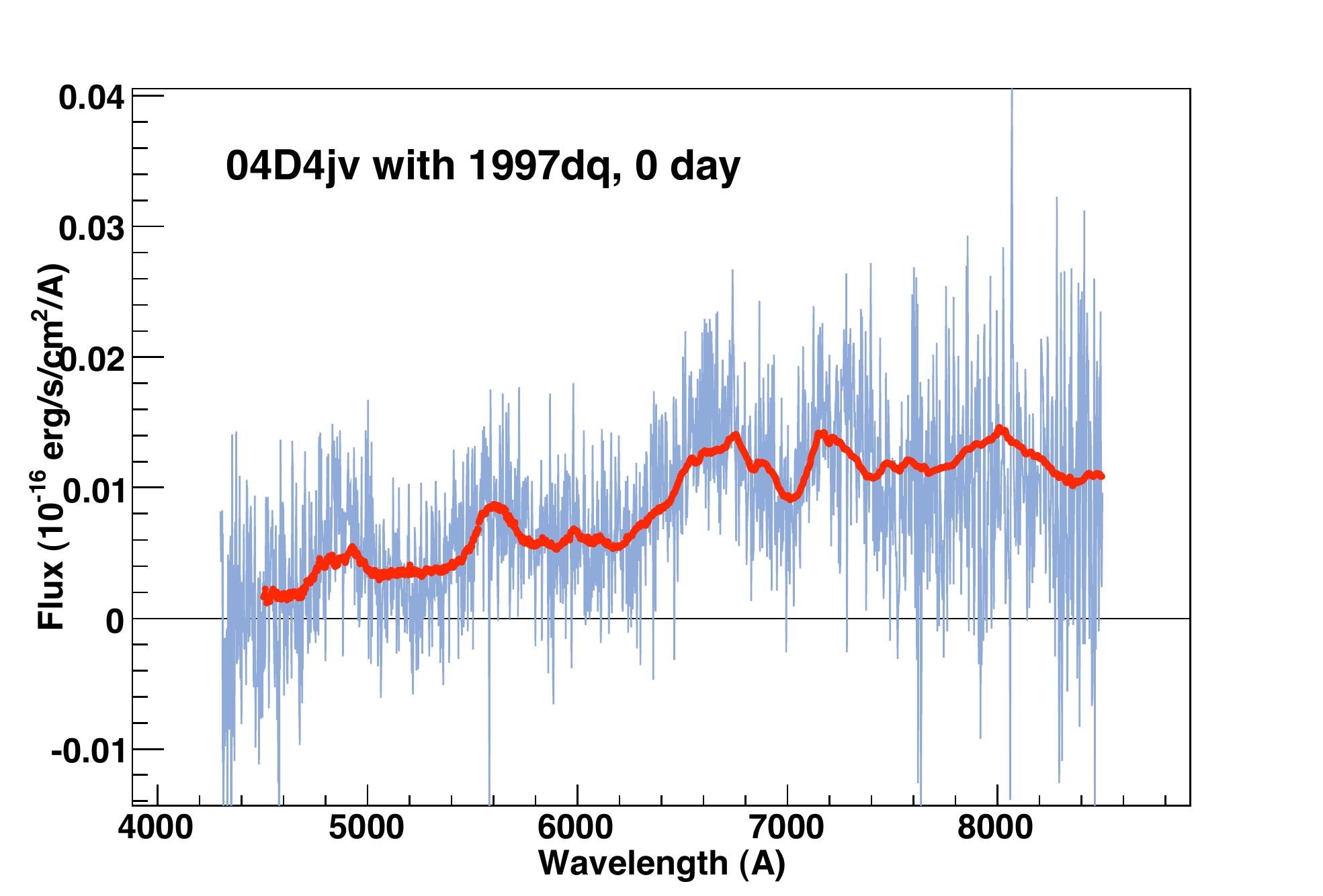}\\
\end{tabular}
\caption{SALT2 fit of SN~04D4jv. Top left: negative deep reference image of SN~04D4jv. The locus of
the SN is indicated as a ``+" sign. The slit position is shown as green dotted lines. Top right: restframe $g_M, r_M, i_M, z_M$ light curves and corresponding
SALT2 fits. Bottom left: PHASE extracted spectrum with SALT2 raw (dashed line) and re-calibrated (solid line) models. Even with
strong re-calibration, SALT2 fails at correctly reproducing the spectrum around 6500 \AA.  Bottom right panel shows a fit of SN~04D4jv with an SN~Ic template (SN~1997dq at maximum light, \citealt{Matheson01}) obtained using the ${\cal SN}$-fit software (see text). 
The spectrum in the right bottom panel is not rebinned. This supernova is identified as an SN~Ic.
  \label{fig:04D4jv}
}
\end{center}
\end{figure*}

The bottom left panel of Fig. \ref{fig:04D4jv} shows the best SALT2 match of 
SN~04D4jv (light curves and spectrum), an SN~Ic at $z=0.228$ around maximum. 
If the light curve fit is visually acceptable, the SALT2 colour index is 
however very high: $c$=0.96 (see Tab. \ref{tab1}; 
by construction the 
average $c$ of the SALT2 training sample is 0). 
SALT2 clearly fails at reproducing the SN structure seen in the 
spectrum around 4800 \AA~ and between 6200 and 7000 \AA, even with strong re-calibration allowed (compare the dashed and solid lines).  A good match with an SN~Ic 
template  (SN~1997dq at maximum light, \citealt{Matheson01}) is obtained using the ${\cal SN}$-fit software (see Section \ref{sec:fors1lss}) and is shown on the bottom right panel of 
Fig. \ref{fig:04D4jv}. This confirms that SN~04D4jv is an SN~Ic.

\subsubsection{SN~05D4ar: an SN~II supernova}

\begin{figure*}[htbp]
\begin{center}

\begin{tabular}[b]{cc}
  \hspace{-5mm}
  \vspace*{1cm}
  \begin{minipage}[b]{7.5cm}
    \centering
    \fbox{\includegraphics[width=6cm,height=2cm]{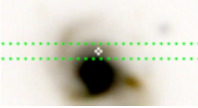}}
  \end{minipage}
  &
  \begin{minipage}[b]{7.5cm}
    \hspace{-5mm}\includegraphics[width=8cm]{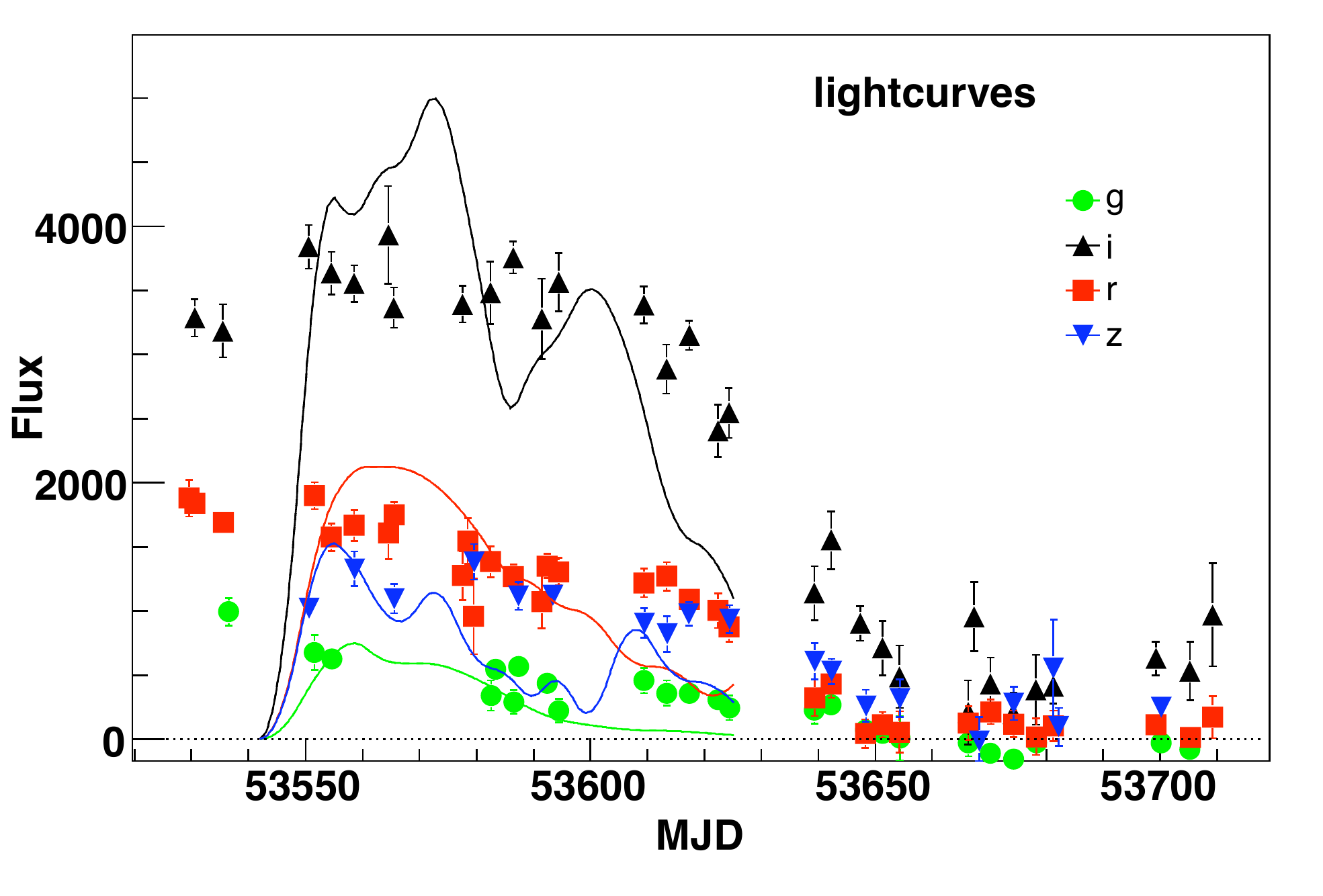}
  \end{minipage}
  \\
  \hspace{-5mm}
  \includegraphics[width=8.5cm]{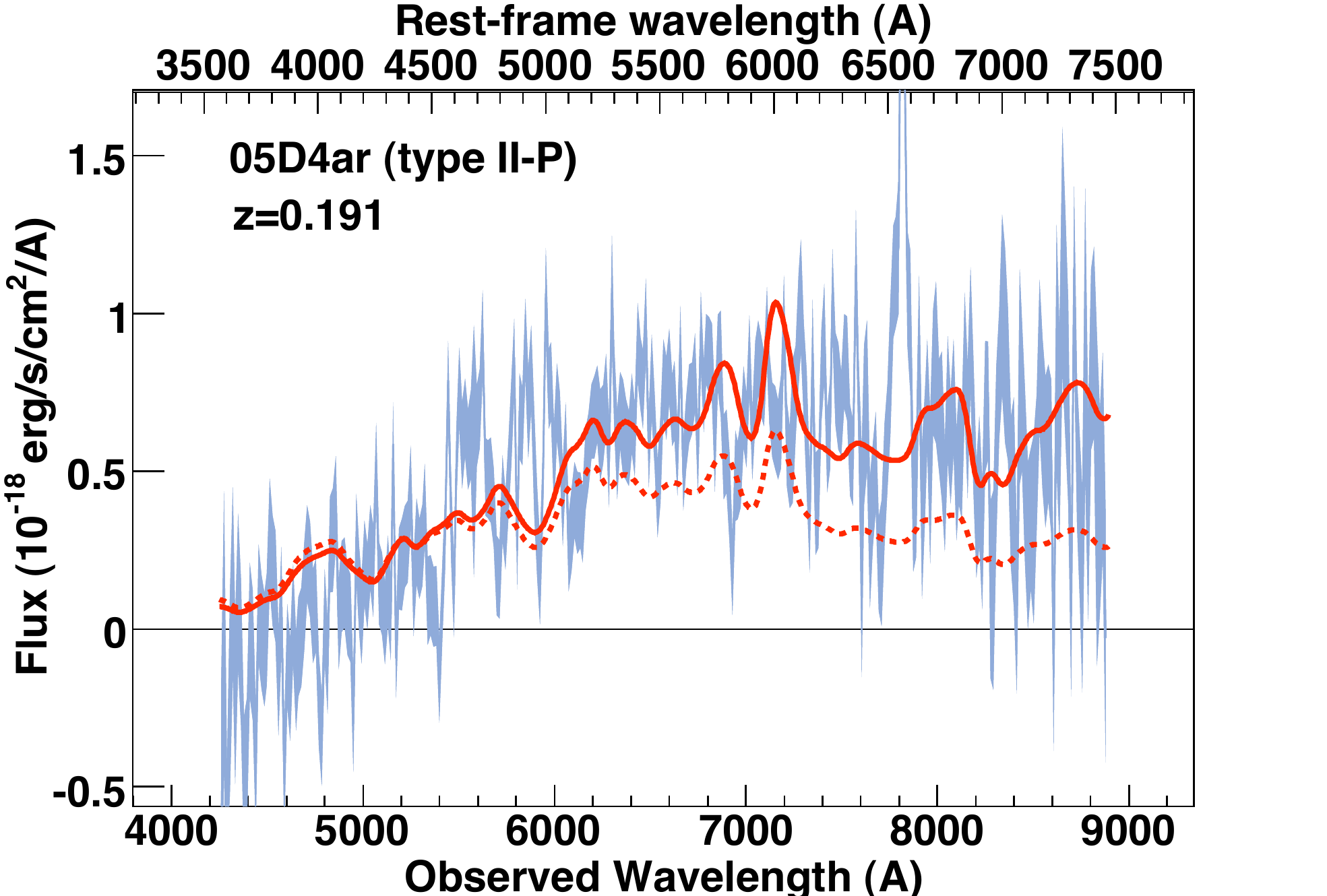}
  &
  \vspace*{-3mm}\includegraphics[width=8.8cm]{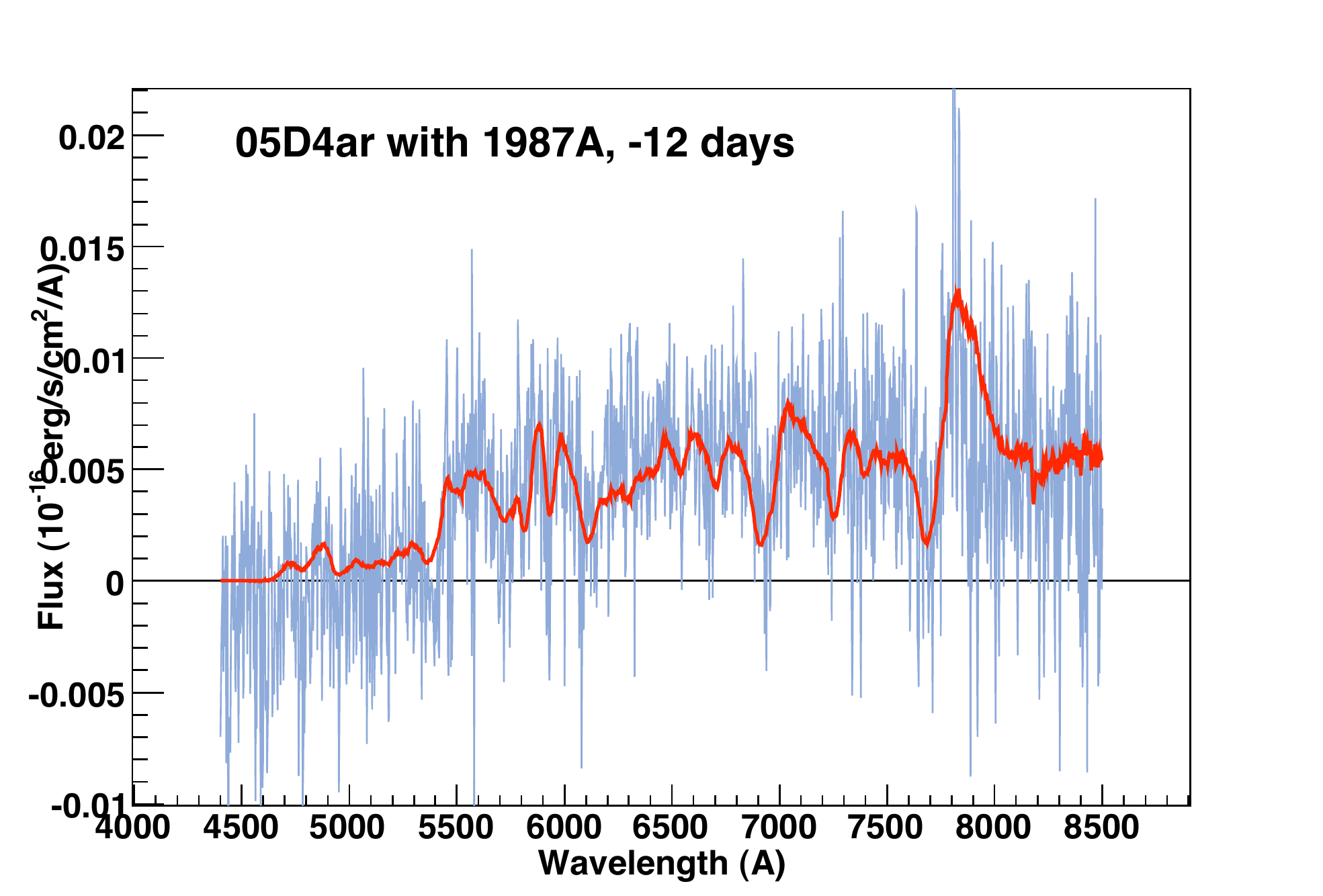}\\
\end{tabular}
\caption{SALT2 fit of SN~05D4ar. Top left: negative deep reference image of SN~05D4ar. The locus of
the SN is indicated as a ``+" sign. The slit position is shown as green dotted lines. Top right: restframe $g_M, r_M, i_M, z_M$ light curves and corresponding
SALT2 fits. SALT2 does not correctly fits these flat light curves. 
Bottom left: PHASE extracted spectrum with SALT2 raw (dashed line) and re-calibrated (solid line) models. Even with
strong re-calibration, SALT2 fails at correctly reproducing the spectrum, namely the $H_\alpha$ P-Cygni feature at 7900 $\AA$ (observer frame). Bottom right panel shows a fit of SN~05D4ar with an SN~II template (SN~1987A, $\phi=-12$ days, \citealt{Pun95}) obtained using the ${\cal SN}$-fit software (see text). The spectrum in the right bottom panel is not rebinned. This supernova is identified as an SN~II.
  \label{fig:05D4ar}
}
\end{center}
\end{figure*}

In a similar way, we identify SN~05D4ar as an SN~II (Fig. \ref{fig:05D4ar}). Here, both the light curve and spectrum SALT2 fits are very poor. Both $x_1$ and $c$ are very discrepant from the average values of the training sample: $x_1=5$ and $c=1.29$, see Tab. \ref{tab1}. The bottom right panel shows the fit obtained with ${\cal SN}$-fit using SN~II templates. The
best-fit is for SN~1987A a few days after explosion (-12 days with respect to maximum light, \citealt{Pun95}). The H$\alpha$ P-Cygni emission is well reproduced, as are 
most of the noticeable features of the spectrum. Combined to the obvious flatness of the light curves, this fact confirms that SN~05D4ar is an SN~II supernova. 

\section{Discussion}

We have taken advantage of SNLS having two independent pipelines 
to cross-check the identifications obtained using SALT2 with the ones done with more standard procedures of template
fitting. 
For a given object,  an independent assessment of the type and the redshift was performed on each side. On one side, the {\em superfit} code
developed by \citet{Howell05} was run on PHASE extracted spectra, using all possible SN type templates. On the other side, SALT2 was systematically run on all PHASE spectra.
Putting our efforts together, we have thus
carefully double-checked the identifications of a sample of about 250 VLT, Gemini and Keck spectra.
For VLT spectra, both extraction with PHASE and
identification with SALT2 considerably improve the type determination. For Gemini
spectra, not currently supported in PHASE, we only test the effect 
of using SALT2 as opposed to standard template fitting techniques. SALT2 
identifications are straightforward and help in many cases to secure the
type of the SN candidate. In some cases, the template fitting technique does
a better job at reproducing the candidate spectrum. This usually happens for
rather peculiar spectra not well represented in the SALT2 training sample. 
A few supernovae that had not been identified previously  and that had been consequently classified as unknown objects, 
have proved to be \Iae in the light of our refined procedure.

Identifications based on the $\chi^2$ fitting of the SN spectrum against
a host and an SN template library \citep{Balland06, Balland07, 
Howell05,Lidman05} improve over the traditional eye-guided
identification of ``typical'' \Iae features. However, they rely on the template library quality that often 
suffers from insufficient phase and wavelength coverage and/or completeness,
especially for non \Iae templates. 
Cross-correlation techniques used by ESSENCE \citep{Matheson05} 
and SDSS \citep{Zheng08}, though sophisticated, suffer to some extent
from the same drawbacks, as they also rely on the comparison to a spectral external template dataset.
\citet{Blondin07} show that the SNID code is able to confidently distinguish an SN~Ic at redshift 0.5 with phase between -5 and + 5 days from other SN types, depending on the use of priors on the redshift and age (see Fig. 21 of \citet{Blondin07} for details). However, this method has the limitation that it can
not subtract host galaxy.

Using SALT2 takes advantage of the spectrophotometric model of \Iae built from a large collection of spectra and light curves of 
local and distant SNe~Ia, a fraction of which being SNLS supernovae. One of the key advantages of SALT2 is
that the model improves as more \Iae are detected and identified as part
of SNLS and external collaboration efforts.  
Moreover, using simultaneous photometric
and spectroscopic information narrows the parameter space available to 
reproduce the SN properties. As an example, with SALT2, the phase is well constrained to 
within a fraction of a day, which allows us to alleviate the possible 
confusion with other  types, such as SNe~Ic. Non \Iae events are detected, as 
they produce parameters values or fits that deviate from the ones of the 
average sample. As a model is compared to the light curves and spectral data, 
the limitation due to incomplete spectral and phase sampling is not as strong 
as in the standard template fitting technique. All supernova candidates are treated on an equal footing and using SALT2 limits the subjective part of the
identification introduced by human judgement. If SALT2
helps at discriminating non \Iae candidates, the current version of the code does not help in identifying
their type. For such cases, recourse to a standard template fitting technique 
is required to secure the type.

\section{Conclusion}

We have developed new techniques for both supernova spectral extraction
and identification. These new tools have been developed for the purpose
of making use of the deep imaging obtained at CFHT to obtain an homogeneous set of SNLS supernova spectra.

Considerable effort has been
put into extracting, in the most efficient way, VLT spectra in order
to get a clean set of supernova spectra at redshifts ranging from $z=0.1$ 
to $z\approx 1$. One key feature of our so-called PHASE extraction is that
the extraction is guided by computing the trace of
the supernova on the spectrogram. This is done by using deep CHTLS reference
images, in various photometric bands, to recover the profile of the 
non-transient objects present in the slit, from which a multi-component model 
is built by adding the (point-like) supernova flux as a Gaussian of width 
equal to the seeing. Fitting the model to the spectrogram light yields the
flux of each component in each pixel. As we have shown, this improves the 
host-supernova separation over more standard techniques. A second key
feature of the PHASE extraction is that it avoids re-sampling the data and
correlating pixels along the procedure. This ensures a clean estimation
of the error level associated with the signal used later for the identification.

We use the spectrophotometric model of SALT2 \citep{Guy07} as an help to determine
the type of supernova candidates. Taking advantage of an ever increasing 
training set as new supernovae are discovered and followed up
by the SNLS and other collaborations, the model is able to adequately
reproduce \Ia features in a wide range of cases (\citealt{Balland08} show that
it is true from phases as
early as -10,-15 days up to a few weeks after maximum light and
for a wide spectral range, including the UV region down to 2100 $\AA$ for the
most distant supernovae). 
The combined fit of light curves and spectrum tightens the constraints
on the parameters that describe the supernova. Non \Iae candidates 
are identified on a case-by-case basis as their spectrophotometric 
best-fitting parameters deviate from the average properties of the SNe~Ia 
training sample. Uncertainty however remains for the most host-contaminated
(host fraction $>$ 70-80\%) and most distant ($z\gesim 1$) cases.

PHASE extractions have been tested against simulations, and SALT2 based identifications have been cross-checked
with template fitting identifications performed on the VLT data using
the technique described in \citet{Howell05}. 

We find that using SALT2 for the purpose of 
identification improves the reliability of the type determination over standard 
techniques. For VLT spectra, PHASE extraction combined to SALT2 
identification takes the most possible out of the data
to secure a clean type and redshift determination. 

\begin{acknowledgements}
We gratefully acknowledge the dedicated work of the day time and night time support staff at the Cerro Paranal Observatory. We also acknowledge the
anonymous referee for helpful comments on the manuscript.
French authors acknowledge support from CNRS/IN2P3,
CNRS/INSU and PNC.

\end{acknowledgements}

\bibliographystyle{aa}
\bibliography{bibi}

\newpage

\newpage

\vfill
\eject

\newpage

\newpage

\newpage

\newpage

\newpage

\newpage

\vfill
\eject

\newpage

\vfill
\eject

\newpage

\vfill
\eject

\newpage

\vfill
\eject

\newpage

\newpage

\newpage
\clearpage

\small
\begin{longtable}[c]{lccc|ccccc}
\caption[]{\label{tab1} Results of SALT2 fits used as an help for identification}\\
\hline\hline 
\\
\multicolumn{4}{c}{SN} & \multicolumn{5}{c}{SALT2 model}\\
\multicolumn{4}{c}{------------------------------------------------------------} & \multicolumn{5}{c}{--------------------------------------------------------}\\
Name  & $z$ & Phase (days/max) & {\bf $d_{SN-Host\ centre}^a$} & $x_1^b$ & $c^b$ & Host Model & \% host$^c$ & $\chi^2$/d.o.f.\\[2mm]
\hline
\\
04D1dc & 0.211 & -0.4 & 1.18 & -1.56 & 0.01 & NoGalaxy$^d$ & 0 & 3.24 \\
05D2dt & 0.574 & -1.7 & 0.01 & 0.01 & 0.08 & E &  43 & 1.13 \\
04D4ib & 0.704 & 0.7 & 0.18 & 0.60 & -0.10 & S0 & 71 & 0.61\\ 
04D4dw & 1.031 & 2.1 & 0.11& 1.35 & 0.02 & Sa & 68 & 1.03 \\
04D4jv & 0.228 & -1 & 1.5& -0.30 & 0.96 & NoGalaxy$^d$ & 0 & 1.43\\
05D4ar & 0.191 & 1.3 & 0.7 & 5.00 & 1.29 & NoGalaxy$^d$ & 0 & 1.71\\[2mm]

\hline
\end{longtable}

\noindent
$^a$ Distance (in arcsec) of the supernova centre to its host centre in the
PHASE extraction model.

\noindent
$^b$ From the SALT2 fit of the light curves.

\noindent
$^c$ Percentage of host signal in the best-fitting model, averaged over
the whole spectral range.

\noindent
$^d$ The best-fitting SALT2 model has no host component.

\end{document}